\documentclass[12pt,twoside]{article}
\usepackage{amssymb}
\usepackage{graphicx}
\usepackage{a4wide}
\begin{document}
{\renewcommand{\thefootnote}{\fnsymbol{footnote}}
\hfill  IGPG--06/2--2\\
\mbox{}\hfill AEI--2006--009\\
\mbox{}\hfill gr--qc/0602100\\[-5mm]
\thispagestyle{empty}
\setcounter{page}{-1}
\begin{center}
{\Huge  Quantum Riemannian Geometry\\[2mm] and Black Holes}\\
\vspace{1.5em}
{\large Martin Bojowald}\footnote{e-mail address: {\tt bojowald@gravity.psu.edu}}\\
\vspace{0.5em}
Institute for Gravitational Physics and Geometry, The Pennsylvania
State University, 104 Davey Lab, University Park, PA 16802, USA;\\
\vspace{0.5em}
Max-Planck-Institut f\"ur Gravitationsphysik, Albert-Einstein-Institut,\\
Am M\"uhlenberg 1, D-14476 Potsdam, Germany
\end{center}
}

\setcounter{footnote}{0}

\newcommand{\Le}{\delta}
\newcommand{\vt}{\vartheta}
\newcommand{\vp}{\varphi}

\newcommand{\case}[2]{{\textstyle \frac{#1}{#2}}}
\newcommand{\lP}{\ell_{\mathrm P}}

\newcommand{\md}{{\mathrm{d}}}
\newcommand{\Kern}{\mathop{\mathrm{ker}}}
\newcommand{\tr}{\mathop{\mathrm{tr}}}
\newcommand{\sgn}{\mathop{\mathrm{sgn}}}

\newcommand*{\R}{{\mathbb R}}
\newcommand*{\N}{{\mathbb N}}
\newcommand*{\Z}{{\mathbb Z}}

\subsection*{Abstract}
  Black Holes have always played a central role in investigations of
  quantum gravity. This includes both conceptual issues such as the
  role of classical singularities and information loss, and technical
  ones to probe the consistency of candidate theories. Lacking a full
  theory of quantum gravity, such studies had long been restricted to
  black hole models which include some aspects of quantization.
  However, it is then not always clear whether the results are
  consequences of quantum gravity per se or of the particular steps
  one had undertaken to bring the system into a treatable form. Over a
  little more than the last decade loop quantum gravity has emerged as
  a widely studied candidate for quantum gravity, where it is now
  possible to introduce black hole models within a quantum theory of
  gravity. This makes it possible to use only quantum effects which
  are known to arise also in the full theory, but still work in a
  rather simple and physically interesting context of black holes.
  Recent developments have now led to the first physical results about
  non-rotating quantum black holes obtained in this way. Restricting
  to the interior inside the Schwarzschild horizon, the resulting
  quantum model is free of the classical singularity, which is a
  consequence of discrete quantum geometry taking over for the
  continuous classical space-time picture. This fact results in a
  change of paradigm concerning the information loss problem.  The
  horizon itself can also be studied in the quantum theory by imposing
  horizon conditions at the level of states. Thereby one can
  illustrate the nature of horizon degrees of freedom and horizon
  fluctuations. All these developments allow us to study the quantum
  dynamics explicitly and in detail which provides a rich ground to
  test the consistency of the full theory.

\newpage
\thispagestyle{empty}
\tableofcontents

\newpage

\section{Introduction}

Black holes in classical general relativity are, compared to other
astrophysical objects, distinguished by the presence of singularities,
where curvature and tidal forces diverge and where space-time stops,
and horizons, which can separate off regions from causal contact from
another region. Both properties have long been suspected to be changed
in a quantum theory of gravity: Singularities denote points where the
classical theory breaks down, and at least space-like ones which lie
to the past or future of observers are supposed to be removed in a
more complete quantum theory. Horizons, on the other hand, are still
expected to play an important role also in quantum gravity. The
horizon surface should at most be smeared out due to fluctuations in
the causal structure on which the concept of horizons relies. For
massive black holes (compared to the Planck mass) these horizon
fluctuations should be negligible for most purposes such that the
classical picture still applies. Instead of modifying the horizon on
large scales, quantum gravity is expected to provide a microscopic
picture which shows how to build a macroscopic horizon from Planck
scale ingredients. If successful, this will then result in a statistical
explanation of black hole entropy.

In more detail, the main issues concerning black holes in quantum
gravity are as follows:
\begin{description}
\item[Singularities:] Are they indeed removed and, if yes, what
  replaces them? There are arguments that not all singularities are
  equal, with space-like ones to be removed and time-like ones to
  persist in order to rule out unwanted (such as negative mass)
  solutions \cite{SingValue}. Also the issue of naked singularities
  and cosmic censorship arises in this context.
\item[Horizons:] First of all, one has to see what an adequate
  definition of a horizon in quantum gravity could be. The original
  concept of the event horizon relies on the classical causal
  structure of all of space-time as well as the presence of
  singularities in the future.  The quasi-local concept of isolated or
  dynamical horizons \cite{HorRev} uses much weaker assumptions about
  the structure of space-time such that it is better suited to a
  quantum treatment at least for large, semiclassical black holes
  which have only weak curvature at the horizon. For microscopic or
  primordial black holes, space-time even around the horizon cannot be
  treated as a smooth classical geometry with a classical causal
  structure. Here it is not clear if a quantum concept of horizon can
  even be applied.
   
   If there is an applicable notion of quantum horizon, the issue of
   black hole entropy can be analyzed. By identifying and counting
   quantum states uniquely characterizing a horizon of a given area one
   can compute black hole entropy and compare with the expected
   semiclassical Bekenstein--Hawking formula. Moreover, detailed
   pictures of the horizon structure and its fluctuations can be
   developed which shed more light on quantum gravity in general. If
   matter fields are present, the horizon should shrink from Hawking
   radiation which provides insights on how gravity interacts with
   matter at the quantum level.
 \item[Both:] Systems such as black holes with singularities as well
   as horizons have led to much confusion in attempts to guess the
   outcome of quantum gravity from early glimpses obtained from mainly
   semiclassical considerations. This is most commonly expressed in
   the infamous information loss paradox according to which
   information falling into the singularity implies a non-unitary
   quantum evolution and thus presumably fundamental limitations to
   knowledge \cite{PredictBreakdown}. These ideas obviously do not
   take into account what happens to singularities in quantum gravity
   and thus have to be revisited once a more complete treatment is
   available.
\end{description}

All these issues probe different aspects of the full theory of quantum
gravity and require different techniques. A common feature, except for
the entropy counting of isolated horizons, is that they are dynamical
aspects such that the Hamiltonian constraint operator in a canonical
quantization or an alternative evolution equation is essential. In
particular, both black hole singularities as well as their horizons
require inhomogeneous situations and an approximation by spatial
homogeneity, which works well in cosmological cases, is not sufficient
in general to grasp all the important physical aspects. This has the
advantage of providing many non-trivial tests of quantum gravity which
go beyond what is possible in homogeneous cosmological models.

It certainly also implies that the treatment is more complicated, and
indeed progress on the problems listed here has been mixed. The
strongest results exist for the counting of black hole entropy of
static or isolated horizons which has been derived in different
approaches
\cite{StringEntro,R:LoopEntro1,K:LoopEntro1,ABCK:LoopEntro,IHEntro}.
This has been possible since the isolation (or even extremality in
\cite{StringEntro}) allows one to ignore the complicated quantum
dynamics and still compute the correct number of physical states.
Moreover, only the horizon itself is important such that its
inhomogeneous neighborhood does not have much influence. This changes
if one also wants to study, e.g., horizon fluctuations since they are
dynamical and require the neighborhood in which the horizon
fluctuates. Thus, both the quantum dynamics and inhomogeneous
configurations have to be handled, and there are not many results
within a full candidate of quantum gravity so far.

Similarly, the issue of singularities relies on dynamical aspects
which for most of the time was too complicated to allow definitive
conclusions as to whether or not singularities persist in quantum
gravity. In the last few years, there has been progress on the
homogeneous situation of cosmological singularities
\cite{IsoCosmo,HomCosmo,Spin} which have been shown to be removed by
quantum gravity \cite{Sing}. Analogous techniques are now also
available for some inhomogeneous situations such as the spherically
symmetric model \cite{SphSymm,SphSymmHam} which is classically
relevant for non-rotating black holes. 
This has led to an extension of the non-singularity statements
from homogeneous models to the spherically symmetric one
\cite{SphSymmSing}.
Moreover, with new results about quantum horizons a consistent picture
of quantum physics of black holes is emerging.

This chapter is also intended as an introduction, by way of examples,
to some of the techniques of quantum geometry with an emphasis on
aspects which are typical for a loop quantization and essential for
physical issues. The main theme will be the understanding of quantum
dynamics in inhomogeneous situations and problems surrounding it.

\section[Spherical Symmetry]{Classical aspects of spherically symmetric systems}

A spherically symmetric metric is most easily written in polar
coordinates $(x,\vt,\vp)$ and takes the form (with $\md\Omega^2=
\md\vt^2+\sin^2\vt\md\vp^2$)
\begin{equation}
 \md s^2 = -N(x,t)^2\md t^2+q_{xx}(x,t)(\md x+N^x(x,t)\md t)^2+
 q_{\vp\vp}(x,t)\md\Omega^2
\end{equation}
where fields only depend on time $t$ and the coordinate $x$ of the
1-dimensional radial manifold $B$.  This expression makes use of the
lapse function $N(x,t)$ and shift vector $N^x(x,t)$ which are
prescribed by the slicing of space-time into spatial constant-$t$
slices: coordinate time translations are generated by the vector field
\begin{equation} \label{ddt}
 \frac{\partial}{\partial t}=N n+ N^x\frac{\partial}{\partial x}
\end{equation}
with the unit vector field $n$ being normal to the slices.  The
spatial metric on those slices is then
\begin{equation} \label{SphSymmMetric}
 \md q^2 = q_{xx}(x,t)\md x^2+q_{\vp\vp}(x,t)\md\Omega^2
\end{equation}
and extrinsic curvature
\begin{equation} \label{Kdef}
 K_{ab}={\textstyle\frac{1}{2}} {\cal L}_n q_{ab}\,,
\end{equation}
which determines the conjugate $\pi^{ab}=-\frac{1}{2}\sqrt{\det q}
(K^{ab}-q^{ab} K^c_c)$ to the metric in a canonical formulation
\cite{ADM}, takes a similar form $K=K_{xx}(x,t)\md
x^2+K_{\vp\vp}(x,t)\md\Omega^2$.

A well-known example is obtained by the spherically symmetric vacuum
solution to Einstein's field equations, the Schwarzschild metric
\cite{Schwarzschild}
\begin{equation}
 \md s^2 = -(1-2M/x)\md t^2+\frac{1}{1-2M/x}\md x^2+x^2\md\Omega^2
\end{equation}
with the mass parameter $M$. It has the following properties: If we
first restrict our attention to larger $x>2M$, the metric is static
since its coefficients do not depend on time and $N^x=0$.  When $x$
becomes large compared to the mass, i.e.\ if we approach the
asymptotic regime far away from the black hole, the metric becomes
asymptotically flat. The black hole region is characterized by the
horizon which appears at $x=2M$ as a coordinate singularity in the
Schwarzschild metric and can be defined in a coordinate independent
manner as the outer boundary of a region where trapped surfaces, i.e.\ 
envelops of light rays which cannot expand outwards to infinity,
occur. If we enter the black hole region through the horizon we notice
that now $t$ becomes a space-like coordinate since the $tt$ component
changes sign. The role of coordinate time is then played by $x$ on
which the metric coefficients depend.  Thus, the interior is not
static, but since the metric components now do not depend on the
spatial coordinate $t$ it is homogeneous (of Kantowski--Sachs form).

\subsection{Metric and triad}

The metric components $q_{\vp\vp}=x^2$, $q_{xx}=(1-2M/x)^{-1}$ and
$N^2=-g_{tt}=1-2M/x$ can be used to characterize the three different
regimes of a massive black hole with mass $M\gg1$: At asymptotic
infinity we have $x\gg 2M\gg 1$ and thus
\[
 q_{\vp\vp}\gg 1 \qquad q_{xx}\sim 1\,.
\]
At the horizon we have $x\sim 2M$ and
\[
 q_{\vp\vp}\gg1 \qquad q_{xx}\gg1
\]
while at the singularity we have $0\sim x\ll 2M$ and
\[
 q_{\vp\vp}\ll1 \qquad |q_{xx}|\ll 1 \qquad N\gg1\,.
\]
In the latter case, $q_{xx}$ is relevant only if we approach the
singularity on slices with $t$ constant which are time-like inside the
horizon. The lapse function, on the other hand, is the relevant metric
component if we approach the singularity on slices which are space-like
inside.

These regimes of metric components can be used for a first glimpse on
how a quantization may deal with the singularity or horizon. From
cosmological models it is known that expressions for, e.g., curvature
components can be modified when they become large, cutting off
classical divergences (in isotropic cosmology they are all inverse
powers of the scale factor \cite{InvScale}, or spin connection
components in anisotropic models \cite{Spin}). Similarly here, some
spin connection components contain information about intrinsic
curvature. Their form can be obtained from the general expression
(see, e.g., \cite{AsFlat})
\begin{equation} \label{GammaGen}
 \Gamma_a^i = -\epsilon^{ijk}e^b_j (\partial_{[a}e_{b]}^k+
 {\textstyle\frac{1}{2}} e_k^ce_a^l\partial_{[c}e_{b]}^l)
\end{equation}
where $e_a^i$ are components of the co-triad (i.e.\
$e_a^ie_b^i=q_{ab}$) and $e^b_j$ of its inverse. In spherical
symmetry, co-triads take a special form just as the metric
(\ref{SphSymmMetric}) does. Since it does not matter how a triad is
rotated, it need not be exactly invariant under the rotation group
acting on space, but it is enough for it to be invariant up to a gauge
rotation. This is realized for co-triads of the form
\begin{eqnarray} \label{SphSymmCotriad}
 e_a^i\tau_i\md x^a &=& e_x(x)\tau_3\md x +(e_1(x)\tau_1+e_2(x)\tau_2)
 \md\vt+ (e_1(x)\tau_2-e_2(x)\tau_1)\sin\vt\md\vp\nonumber\\
 &=:& e_x(x)\tau_3\md x+e_{\vp}(x)\bar{\Lambda}(x) \md\vt+
 e_{\vp}(x)\Lambda(x)\sin\vt\md\vp
\end{eqnarray}
where we use SU(2) generators $\tau_j=-\frac{i}{2}\sigma_j$ with Pauli
matrices $\sigma_j$, and $\Lambda=:\cos\eta \tau_2+\sin\eta \tau_1$
and $\bar{\Lambda}:= \exp(-\frac{\pi}{2}\tau_3) \Lambda
\exp(\frac{\pi}{2}\tau_3)$ are defined to have unit norm in su(2),
i.e.\ $\cos\eta=e_1/e_{\vp}$ and $\sin\eta=-e_2/e_{\vp}$ with
$e_{\vp}^2=e_1^2+e_2^2$. Infinitesimal rotations of space now act by
Lie derivatives on $e$ with respect to superpositions of vector fields
$X=\sin\vp\partial_{\vt}+\cot\vt\cos\vp\partial_{\vp}$,
$Y=-\cos\vp\partial_{\vt}+\cot\vt\sin\vp\partial_{\vp}$, and
$Z=\partial_{\vp}$, while gauge rotations of the triad act by
conjugation in su(2). We thus obtain explicitly
\begin{eqnarray*}
 {\cal L}_Xe &=& (e_1\tau_1+e_2\tau_2)\cos\vp\md\vp-
 (-e_2\tau_1+e_1\tau_2)\frac{\cos\vp}{\sin\vt}\md\vt
 = \left[e,\frac{\cos\vp}{\sin\vt}\tau_3\right]\\
 {\cal L}_Ye &=& (e_1\tau_1+e_2\tau_2)\sin\vp\md\vp-
 (-e_2\tau_1+e_1\tau_2)\frac{\sin\vp}{\sin\vt}\md\vt
 = \left[e,\frac{\sin\vp}{\sin\vt}\tau_3\right]\\
{\cal L}_Ze &=& 0
\end{eqnarray*}
showing that any rotation in space simply amounts to a gauge rotation
of the triad. The corresponding metric is thus invariant under
rotations, and indeed a co-triad (\ref{SphSymmCotriad}) implies a
metric of the form (\ref{SphSymmMetric}) with
\begin{equation}
 q_{xx}=e_x^2 \quad,\quad q_{\vp\vp}=e_1^2+e_2^2=e_{\vp}^2\,.
\end{equation}

A spherically symmetric spin connection takes the form
\begin{equation} \label{SphSymmSpin}
 \Gamma_a^i\tau_i\md x^a = \Gamma_x\tau_3\md x+
 \Gamma_{\vp}\bar{\Lambda}^{\Gamma}\md\vt+
 \Gamma_{\vp}\Lambda^{\Gamma}\sin\vt\md\vp+ \tau_3\cos\vt\md\vp
\end{equation}
where the last term must be added since a connection transforms
differently from a co-triad under gauge transformations. Indeed,
\begin{eqnarray*}
 {\cal L}_X(\tau_3\cos\vt\md\vp) &=& - \tau_3
 \left(\frac{\sin\vp}{\sin\vt}\md\vp+
 \frac{\cos\vt\cos\vp}{\sin^2\vt}\md\vt\right)=
\md\left(\frac{\cos\vp}{\sin\vt}\tau_3\right)\\
{\cal L}_Y(\tau_3\cos\vt\md\vp) &=& -
 \tau_3\left(\frac{\cos\vp}{\sin\vt}\md\vp-
 \frac{\cos\vt\sin\vp}{\sin^2\vt}\md\vt\right)
 = \md\left(\frac{\sin\vp}{\sin\vt}\tau_3\right)
\end{eqnarray*}
and ${\cal L}_Z(\tau_3\cos\vt\md\vp)=0$ such that we have the correct
transformation of $\Gamma$ with the same gauge rotation as above.

The explicit formula (\ref{GammaGen}) applied to a spherically
symmetric co-triad shows that
\begin{equation} \label{GammaComp}
 \Gamma_x=-\eta'\quad,\quad \Gamma_{\vp}=-e_{\vp}'/e_x \quad,\quad
 \Lambda^{\Gamma}= \bar\Lambda\,,
\end{equation}
with $\bar\Lambda$ as defined for the co-triad, such that the
$\vp$-component $\Gamma_{\vp}$ is gauge invariant while $\Gamma_x$ is
pure gauge. Modifications to classical behavior similar to those in
cosmological models can now be expected, e.g., from the spin
connection component $\Gamma_{\vp}=-\sqrt{q_{\vp\vp}}'/\sqrt{q_{xx}}$
when metric components become small. Classically, this expression
diverges at small $q_{xx}$, which can be changed in a quantum theory
for the corresponding operator. Since, as we will see later,
$\Gamma_{\vp}$ appears in the equations of motion, a modification here
would change the behavior of solutions. Horizons of massive black
holes would remain unmodified since there both metric components are
large. The singularity, however, looks less clear: only for time-like
slices does $q_{xx}$ become small, indicating a modification and the
possibility of removal of the singularity. But if we approach the
singularity on space-like slices with $x$ constant, in the interior
$N^2$ (playing then the role of $q_{xx}$) remains large which does not
suggest modifications. Indeed, the slices then are homogeneous and
$\Gamma_{\vp}$ vanishes identically which means that we will need
another measure for the removal of singularities in this case.

At asymptotic infinity, however, we would encounter severe problems
since $q_{xx}$ is close to one at which point modifications can
already be noticeable, spoiling the classical limit of the theory.
This is a sign of using the wrong variables since the modification is
a consequence of quantum effects, and the success of a quantization
can depend significantly on the choice of fundamental variables.
Indeed, there are variables better suited to a demarkation of the
different regimes than the metric. This is in particular the case for
the densitized triad defined by $E^a_i=e^a_i |\det (e_b^j)|$ where
$e^a_i$ is the inverse of the co-triad $e$ compatible with the metric.
A spherically symmetric densitized triad is of the general form
\begin{equation} \label{E}
 E=E^x(x)\tau_3\sin\vt\frac{\partial}{\partial x}+
(E^1(x)\tau_1+E^2(x)\tau_2)\sin\vt\frac{\partial}{\partial\vt}+
(E^1(x)\tau_2-E^2(x)\tau_1)\frac{\partial}{\partial\vp}
\end{equation}
written down as an su(2) valued densitized vector field.  The gauge
invariant components are $E^x$ and $(E^{\vp})^2= (E^1)^2+(E^2)^2$
whose relation with the metric components is
\begin{equation}
 |E^x|=q_{\vp\vp} \qquad E^{\vp}=\sqrt{q_{xx}q_{\vp\vp}}
\end{equation}
(note that $E^x$ can be positive or negative depending on the
orientation $\sgn\det E=\sgn E^x(E^{\vp})^2$ of the triad). The
angular components have the same internal directions $\Lambda$ and
$\bar\Lambda$ as the co-triad.

For the Schwarzschild solution with $|E^x|=x^2$ and
$E^{\vp}=x/\sqrt{1-2M/x}$ we now have the following behavior: At
asymptotic infinity
\[
 |E^x|\gg 1 \qquad E^{\vp}\gg1\,,
\]
at the horizon
\[
 |E^x|\gg1 \qquad E^{\vp}\gg1
\]
and at the singularity
\[
 |E^x|\ll1 \qquad E^{\vp}\ll 1\,.
\]
Thus, irrespective of the approach to the singularity, the behavior is
just as needed for unmodified classical behavior far away from the
black hole all the way up to the horizon, while inverse triad
components, such as the spin connection component
\begin{equation} \label{Gammap}
 \Gamma_{\vp}=-(E^x)'/2E^{\vp}
\end{equation}
will be modified at the singularity with small $E^{\vp}$.

\subsection{Basic variables}

For detecting the classical singularity it seems much more reliable to
use the densitized triad rather than the metric, which is also the
case in homogeneous models with an explicit impact on the removal of
singularities \cite{HomCosmo}. Indeed, the densitized triad as a basic
variable is important in other ways, too: it arises naturally when one
attempts to quantize gravity in a background independent manner. These
two issues, the fate of classical singularities and background
independence, are superficially quite different but turn out to be
deeply related.

Most recent progress in a background independent quantization of
general relativity has come after a reformulation in terms of Ashtekar
variables \cite{AshVar,AshVarReell} where the densitized triad $E^a_i$
plays the role of a momentum canonically conjugate to the Ashtekar
connection $A_a^i=\Gamma_a^i-\gamma K_a^i$ with the spin connection
$\Gamma_a^i$ as a function of $E^a_i$ via (\ref{GammaGen}), extrinsic
curvature $K_a^i= e^b_i K_{ab}$ and the Barbero--Immirzi parameter
$\gamma>0$ \cite{Immirzi}. The extrinsic curvature components here
make $A_a^i$ canonically conjugate to $E^a_i$, while the spin
connection provides $A_a^i$ with the transformation properties of a
connection. This reformulation thus casts general relativity as a
gauge theory and does not only bring it formally closer to other
interactions but also leads to a direct way for a background
independent quantization.

Usually, a field theory would be quantized by smearing the fields with
test functions over 3-dimensional regions so as to make their
classical Poisson *-algebra well defined. For instance, a scalar
$\phi$ with Lagrangian $\sqrt{\det q}(\frac{1}{2} \dot{\phi}^2+
\frac{1}{2} q^{ab}\partial_a\phi\partial_b\phi+ V(\phi))$ on a
background metric $q_{ab}$ (assuming lapse function $N=1$ and shift
vector $N^a=0$) has momentum $p_{\phi}=\sqrt{\det q}\dot{\phi}$ which
transforms as a density (which is often ignored when the background
metric is fixed as, e.g., Minkowski space). This has the singular
Poisson relations $\{\phi(x),p_{\phi}(y)\}=\delta(x,y)$. However, if
we smear the fields with test functions $f$ and $g$ on space to obtain
$\phi[f]:=\int\sqrt{\det q}f(x)\phi(x)\md^3x$ and $p_{\phi}[g]:=\int
g(x)p_{\phi}(x)\md^3x$ we obtain the well-defined Poisson algebra
$\{\phi[f],p_{\phi}[g]\}= \int\sqrt{\det q}f(x)g(x)\md^3x$. This does
not contain $\delta$-functions, but does depend on the background
metric $q$ which is not available for a background independent
formulation of gravity. The very first step of a background independent
quantization of general relativity, therefore, has to face the problem
that the physical fields, with the metric or densitized triad among
them, need to be smeared for a well-defined algebra to be represented
on a Hilbert space, but that a background metric must not be
introduced.

For a scalar, there is a simple way out: as is easily verified, we
still obtain a well-defined algebra if we only smear $p_{\phi}$ for
which we do not need a background metric since it is already a
density.  Similarly, in the case of gravity we can evade the problem
in Ashtekar's formulation since with connections and densitized vector
fields as basic variables there is a natural, background independent
smearing leading to a well-defined algebra: Instead of 3-dimensional
smearings for all basic fields we use a 1-dimensional smearing of the
connection and a 2-dimensional one for the densitized triad, giving
rise to holonomies
\begin{equation}
 h_e(A)={\cal P}\exp\int_e\tau_i A_a^i\dot{e}^a{\rm d}t
\end{equation}
along edges $e$ in space, and fluxes
\begin{equation} \label{Flux}
 F_S(E)=\int_S \tau^i E^a_in_a{\rm d}^2y 
\end{equation}
through surfaces $S$. (We use the tangent vector $\dot{e}^a$ to the
curve $e$ and the co-normal $n_a$ to the surface $S$, both of which
are defined without reference to a metric.)

It turns out that this smearing is sufficient for a well-defined
classical Poisson algebra which even has a unique diffeomorphism
invariant representation
\cite{FluxAlg,Meas,HolFluxRep,SuperSel,WeylRep}. This representation
defines the basic framework of loop quantum gravity
\cite{Rov:Loops,ThomasRev,ALRev,Rov}. States are represented usually in
the connection representation $\psi[A]$ on which holonomies act as
multiplication operators and fluxes as derivative operators. This can
all be done rigorously thanks to a rich structure on the infinite
dimensional space of connections which is under much better control
than the space of metrics. As a consequence, flux operators have
discrete spectra implying a discrete structure of spatial geometry
\cite{AreaVol,Area,Vol2} which is also realized in symmetric models
\cite{cosmoII,SphSymmVol}. Moreover, since flux spectra are discrete
and contain zero, there are no densely defined inverse operators.
Instead there are techniques \cite{QSDV} which allow one to quantize
co-triad or other inverse components of the basic $E^a_i$ by operators
which reduce to the inverse in a classical regime but modify the
classical divergence at small values.  This has already been described
and used above for the spherically symmetric spin connection
component. Here, it is important that those expressions are taken as
functions of the densitized triad components and not metric
components. These effects come from properties of flux operators as
basic operators in a background independent formulation which relies
on the densitized triad as basic variable and so far is not known in a
metric formulation. Indeed, as observed before, the densitized triad is
much better suited to separate the classical singularity from other
regimes such that modifications are only expected there.

\subsection{Dynamics}

Up until now we have discussed kinematical properties of the spherically
symmetric system. The dynamical behavior of triad and connection (or
extrinsic curvature) components is dictated by the Hamiltonian
constraint
\begin{equation} \label{HSphSymm}
 H[N] = (2G)^{-1}\int_B\md x N(x) |E^x|^{-1/2}\left(
 (K_{\vp}^2 E^{\vp}+2
 K_{\vp}K_x E^x)+(1-\Gamma_{\vp}^2)E^{\vp}+ 2\Gamma_{\vp}' E^x \right)
\end{equation}
in terms of the spin connection component $\Gamma_{\vp}$ as before and
the extrinsic curvature components in
\begin{equation} \label{K}
 K=K_x(x)\tau_3\md x+(K_1(x)\tau_1+K_2(x)\tau_2)\md\vt+
(K_1(x)\tau_2-K_2(x)\tau_1)\sin\vt\md\vp
\end{equation}
where again only $K_x$ and $K_{\vp}^2=K_1^2+K_2^2$ are gauge
invariant. In addition, there is the diffeomorphism constraint
\begin{equation} \label{DSphSymm}
 D[N^x]=(2G)^{-1} \int_B N^x(x)
 (-2E^{\vp}K_{\vp}'+K_xE^{x\prime})\,.
\end{equation}
Physical fields $(K_x, E^x;K_{\vp},E^{\vp})$ have to solve the
constraint equations $H[N]=0=D[N^x]$ for all functions $N$ and $N^x$
on $B$ (except for possible boundary conditions which we ignore here)
and evolve in coordinate time according to Hamiltonian equations of
motion $\dot{E}^x =
 \{E^x,H[N]+D[N^x]\}$, etc.\ to be computed
with the Poisson relations $\{K_x(x_1),E^x(x_2)\}=-2G\delta(x_1,x_2)$,
$\{K_{\vp}(x_1),E^{\vp}(x_2)=-G\delta(x_1,x_2)$. For the triad
components this gives
\begin{eqnarray}
 \dot{E}^x &=& 2NK_{\vp}\sqrt{|E^x|}+N^xE^{x\prime}
  \label{Er}\\
 \dot{E}^{\vp}  &=& N(K_{\vp}E^{\vp}+K_xE^x)
 |E^x|^{-1/2} +(N^xE^{\vp})' \label{Ep}
\end{eqnarray}
which, when solved for the extrinsic curvature components, agrees with
their geometrical definition via
\begin{equation} \label{ExtrCurv}
 K_a^i=e^b_i K_{ab}= (2N)^{-1}e^b_i {\cal  L}_{\partial_t- N^x
   \partial_x} e^j_ae^j_b
\end{equation}
from (\ref{Kdef}) and (\ref{ddt}).  Evaluating this for a spherically
symmetric co-triad (\ref{SphSymmCotriad}) or densitized triad
(\ref{E}) indeed gives spherically symmetric components
\[
 K_x = N^{-1}(\dot{e}_x-(N^xe_x)')\quad, \quad
 K_{\vp}=N^{-1}(\dot{e}_{\vp}-N^xe_{\vp}')
\]
and the same internal directions $\Lambda^K=\Lambda$,
$\bar\Lambda^K=\bar\Lambda$ as those of the triad. The extrinsic
curvature components then have Hamiltonian equations of motion
\begin{eqnarray}
  \dot{K}_x &=& -NK_{\vp}K_x|E^x|^{-1/2}+\case{1}{2}
NK_{\vp}^2E^{\vp}|E^x|^{-3/2}+(N^xK_x)'\nonumber\\
&&+\case{1}{2}N|E^x|^{-1/2}\left(E^{\vp}|E^x|^{-1}- {\textstyle\frac{1}{4}}
  (E^{x\prime})^2 (|E^x|E^{\vp})^{-1}- E^{x\prime}E^{\vp\prime}
  (E^{\vp})^{-2}+ E^{x\prime\prime}(E^{\vp})^{-1}\right)\nonumber\\
&&+\case{1}{2} N'\left(E^{x\prime}(E^{\vp})^{-1}|E^x|^{-1/2}-
  2\sqrt{|E^x|}E^{\vp\prime}(E^{\vp})^{-2}\right)+
N''\sqrt{|E^x|}(E^{\vp})^{-1}
  \label{Kr}\\
 \dot{K}_{\vp} &=& -\case{1}{2}N K_{\vp}^2|E^x|^{-1/2}+N^xK_{\vp}' \nonumber\\
 &&+\case{1}{2}N|E^x|^{-1/2}(\case{1}{4}(E^{x\prime})^2(E^{\vp})^{-2}-1)
 +\case{1}{2}N'\sqrt{|E^x|}E^{x\prime}(E^{\vp})^{-2} \,. \label{Kp}
\end{eqnarray}

These coupled non-linear equations are difficult to solve in general,
but the Schwarzschild solution can easily be reproduced by assuming
staticity: $K_x=K_{\vp}=N^x=0$ which already implies that the
diffeomorphism constraint is satisfied. Equations (\ref{Kr}) and (\ref{Kp})
then assume the form of consistency conditions for the lapse function
in order to ensure the existence of a static slicing. Both conditions
are identically satisfied for a lapse function 
\begin{equation} \label{Nstatic}
 N\propto E^{x\prime}/E^{\vp}
\end{equation}
using that $E^{\vp}$ and $E^x$ are subject to the constraint equation
\[
 (\Gamma_{\vp}^2-1)E^{\vp}-2\Gamma_{\vp}'E^x=
 (\case{1}{4}(E^{x\prime})^2/(E^{\vp})^2-1)
 E^{\vp}+(E^{x\prime}/E^{\vp})' E^x=0
\]
following from (\ref{HSphSymm}) with $K_x=0=K_{\vp}$ and
$\Gamma_{\vp}$ from (\ref{Gammap}).

It remains to solve this constraint for $E^x$ and $E^{\vp}$.  If we
choose our radial coordinate such that $|E^x|=x^2$, this simplifies to
a differential equation
\[
 -2x^3 E^{\vp\prime}+3x^2E^{\vp}-(E^{\vp})^3=0
\]
whose solution $E^{\vp}(x)=x(1+c/x)^{-1/2}$ is the Schwarzschild
component for $E^{\vp}$ with $c=-2M$, which then also reproduces the
correct lapse function from (\ref{Nstatic}).

This shows how the dynamical equations appear in a canonical
formalism, and also how special the simplicity of the static
Schwarzschild solution is. With slight modifications to the equations,
e.g.\ coming from quantum modifications, the assumption of staticity
will no longer be consistent since two conditions
$\dot{K}_x=0=\dot{K}_{\vp}$ have to be satisfied by only one function
$N$. Thus, quantum corrections are expected to change the static
behavior of the classical solution, even though it would come from
only small changes outside the horizons of massive black holes. What
this means for the inside where quantum effects dominate around the
singularity has to be analyzed by direct methods from quantum gravity.

\section{Quantization: Overview}

Even though the vacuum spherically symmetric system has only a finite
number of physical degrees of freedom given by the black hole ADM mass
and its conjugate momentum \cite{SphKl1,SphKl2,Kuchar}, a Dirac
quantization requires field theory techniques in order to deal with
infinitely many kinematical degrees of freedom. Almost all of these
degrees of freedom will then be removed by the Hamiltonian constraint
which acts as a functional differential or difference operator. Thus,
many of the field theoretic aspects of the full theory can be probed
here which also implies a corresponding level of complication. So far,
the system is not fully understood in a loop quantization even in the
vacuum case, and other techniques which can be applied more easily to
this system do not allow definitive conclusions about the singularity.
It is therefore necessary at this stage to refer to approximation
methods.  These methods allow different glimpses which one can then
try to bring together for a consistent picture. Here, we briefly
collect different classes of approximations, which will be described
in more detail in the following sections.

\subsection{Homogeneous techniques}

Currently, loop techniques for homogeneous geometries, following
techniques introduced in \cite{SymmRed}, are fully developed to a
degree that one can analyze properties of physical solutions. (The
main open issue is the physical inner product, about which not much is
known even in the simplest cases
\cite{Golam,IsoSpinFoam}.) There are explicit expressions for the most
important operators such as the volume operator \cite{cosmoII}, matter
Hamiltonians or the Hamiltonian constraint
\cite{IsoCosmo,HomCosmo,Spin} which is a big advantage compared to the
full theory where the ubiquitous volume operator cannot be
diagonalized even in principle. The constraint equation takes the form
of a difference equation for the wave function in the triad
representation which explicitly shows how one can evolve through the
classical singularity. Moreover, one can define effective classical
equations with diverse correction terms
\cite{Inflation,Time,EffAc,Perturb,Josh,EffHam,DiscCorr}. They capture the
main quantum effects
\cite{Inflation,GenericInfl,BounceClosed,BounceQualitative,GenericBounce,Oscill,Cyclic,InflOsc,JimDual}
and can be analyzed more easily than the quantum difference equation
directly (see, e.g.,
\cite{DynIn,Scalar,ScalarLorentz,ClosedExp,FundamentalDisc,GenFunc,GenFuncBI}).

In some cases these effective classical equations provide an intuitive
explanation for the removal of singularities since they display
bouncing behavior of a cosmological solution. This can also be used to
model the case of matter collapsing into a black hole. As a model, the
ball of matter can be assumed to be homogeneous such that the collapse
of the outer shell radius is described by effective equations for an
isotropic system. These equations are modified at small scales, i.e.\ 
when the ball collapses to a certain size. In the modified regime
there are matter systems which show a bounce, which now can be
interpreted as the collapsing matter parts repelling each other and
bouncing back after maximal contraction. So far, this is not much
different from a bouncing universe and indeed described by the same
equations.  The difference is that the matter ball does not present
the full system, but that there is also the outside. Without
specifying the matter content there, one can try to match the interior
to a generalized Vaidya metric outside allowing for matter radiated
away.  This allows to study the formation or disappearance of horizons
which may or may not shield the bounce replacing the classical
singularity \cite{Collapse}.

Limitations of these techniques are that only the interior carries
quantum effects, while the outside is described by a generalized
Vaidya metric of general relativity. Some quantum effects are
transported to the outside by matching to the effective interior,
which then enter the Vaidya solution effectively through a
non-standard energy momentum tensor. This still shows possible changes
in the behavior of horizons, but is of course more indirect than a
complete inhomogeneous analysis.

A different approach using homogeneous techniques only applies to the
Schwarzschild solution which is homogeneous inside the horizon. One
can then describe the interior by a quantum equation which as in
cosmological cases, is a difference equation not breaking down at the
classical singularity. Also here we thus obtain a mechanism to evolve
through a classical singularity, and there are many more non-trivial
aspects which only arise in a loop quantization and show its
consistency \cite{BHInt}. In particular, the singularity is removed,
but the horizon which presents another boundary to the classical
interior remains.

\subsection{Extrapolation}

The previous analysis provides a picture of a non-zero Schwarzschild
black hole interior which one can now extrapolate in two ways: The
non-singular interior first has to be embedded in a full space-time
which can happen in several different ways. Moreover, for a realistic
black hole this must be generalized to the presence of matter.
While there are many gaps to be filled in by detailed constructions
and calculations, one can already see different implications for the
issue of information loss \cite{BHPara}.

\subsection{Inhomogeneous techniques}

Operators for the spherically symmetric system (with or without
matter) are now available explicitly at a level similar to that in
homogeneous models \cite{SphSymm,SphSymmHam}. In particular, there is
a similar simplification in the volume operator which translates to
matrix elements of the Hamiltonian constraint also being known
explicitly. However, the constraint is much more difficult to analyze
since it now presents a functional difference equation in infinitely
many kinematical variables. The construction and regularization of the
constraint is more subtle compared to homogeneous cases, but
similar to the full theory where there are different versions. These
can then be studied explicitly and their physical implications
analyzed, leading possibly to conclusions as to which operator is most
suited for the full theory.

Even though the singularity issue is not yet solved in generality,
there are indications that a mechanism similar to that in homogeneous
models is at work. This will then provide a large class of systems
where one and the same mechanism, derived from basic loop properties,
provides a removal of singularities in non-trivial ways.

There are regimes where the constraint operator can be approximated by
a simpler expression. Interestingly, this is true in particular in the
neighborhood of isolated or slowly evolving horizons \cite{Horizon}
such that horizon properties such as fluctuations and its growth from
infalling matter or shrinking from Hawking radiation can be analyzed.
Moreover, the regime provides perturbation techniques which allow us to
study general properties of the constraint operator and matter
Hamiltonians.

\subsection{Full theory}

The full theory has a rigorous quantum representation \cite{ALMMT} and
well-defined candidates for the Hamiltonian constraint \cite{QSDI}.
Understanding the dynamics in general is certainly very complicated,
and even computing matrix elements of the constraint is involved. Most
full results which contribute to the physical picture are thus
non-dynamical: Spatial geometry is discrete \cite{AreaVol,Area,Vol2}
as a characteristic of the full quantum representation, and there are
well-defined quantum matter Hamiltonians \cite{QSDV}. Black hole (and
cosmological) horizons can be introduced as a boundary provided they
are isolated \cite{HorRev}. This condition ensures that the dynamics
at the boundary is not essential and allows the correct counting of
black hole entropy \cite{ABCK:LoopEntro,IHEntro}.

\section{Homogeneous techniques}

In the Schwarzschild interior $r<2M$ one can choose a homogeneous slicing
such that the metric is of the Kantowski--Sachs form
\begin{equation}
 \md s^2 = -N(T)^2\md T^2+(2M/T-1)\md R^2+T^2\md\Omega^2
\end{equation}
with $T=r$, $R=t$ and a lapse function $N(T)^2=T/(2M-T)$. The spatial
metric is then related to a homogeneous triad of the form (\ref{E})
where $E^x$ and $E^{\vp}$ are constants on spatial slices. Their
conjugates are given by Ashtekar connection components of the general
spherically symmetric and homogeneous form
\begin{equation} \label{ASphSymm}
 A_a^i\tau_i\md x^a= A_x\tau_3\md x+A_{\vp}\bar\Lambda^A\md\vt+
 A_{\vp}\Lambda^A \sin\vt\md\vp+ \tau_3\cos\vt\md\vp
\end{equation}
which in this case are simply proportional to extrinsic curvature
components $K_x=-A_x/\gamma$ and $K_{\vp}=-A_{\vp}/\gamma$ since
$\Gamma=\tau_3\cos\vt\md\vp$ from (\ref{SphSymmSpin}) and
(\ref{GammaComp}) with homogeneity. Moreover, $\Lambda^A=\Lambda$ (as
defined for the triad) follows from the Gauss constraint. Since
$\Lambda$ is constant in a homogeneous model and subject to gauge
rotations, we will fix it to $\Lambda=\tau_2$ in this section, such
that $\bar\Lambda=\tau_1$.  The symplectic structure for the
4-dimensional phase space is determined by $\{K_x,E^x\}=-2G$,
$\{K_{\vp},E^{\vp}\}=-G$.

\subsection{Quantum representation}

Loop quantum gravity is based on spin network states which are
generated by holonomies as multiplication operators. Similarly,
homogeneous models in loop quantum cosmology are based on a
representation \cite{Bohr} which emerges from holonomies of
homogeneous connections and which turns out to be inequivalent to the
usual Schr\"odinger representation used in a Wheeler--DeWitt like
quantization. For the Kantowski--Sachs model an orthonormal basis of
states is given by the family
\begin{equation} \label{HomBas}
 \langle K_{\vp},K_x|\mu,\nu\rangle = \exp(-\case{i}{2}\gamma(\mu K_{\vp}+\nu
 K_x)) \qquad \mu,\nu\in\R, \mu\geq 0
\end{equation}
such that the kinematical Hilbert space is non-separable. (There are
arguments to reduce this to a separable Hilbert space as in
\cite{cosmoI} using properties of observables \cite{Velhinho}.) One
can see one of the basic loop properties that only exponentials of
connection or extrinsic curvature components are represented directly,
but not the components themselves: It is clear that, e.g.,
$\exp(-i\gamma K_x/2)$ acts directly as a shift operator
\begin{equation}
 \widehat{\exp(-i\gamma \kappa K_x)}|\mu,\nu\rangle=|\mu,\nu+2\kappa\rangle
\end{equation}
but since this operator family is not represented continuously, this
does not allow us to obtain an operator for $K_x$ by
differentiation. Indeed,
\[
 \langle \mu,\nu| \widehat{\exp(-i\gamma \kappa K_x)}|\mu,\nu\rangle=
 \langle\mu,\nu|\mu,\nu+2\kappa\rangle = \delta_{0,\kappa}
\]
is not continuous at $\kappa=0$. This is different from a
Wheeler--DeWitt quantization where extrinsic curvature components
would be basic operators represented directly. Instead, here we have
to express those components through holonomies such as $\exp (\gamma
K_x\tau_3)=\cos (\frac{1}{2}\gamma K_x)+2\tau_3\sin (\frac{1}{2}\gamma
K_x)$ and use the action
\begin{eqnarray}
 \cos (\case{1}{2}\gamma K_x)|\mu,\nu\rangle &=&
 \case{1}{2}(|\mu,\nu+1\rangle+|\mu,\nu-1\rangle)\\
 \sin (\case{1}{2}\gamma K_x)|\mu,\nu\rangle &=&
 \case{i}{2}(|\mu,\nu-1\rangle-|\mu,\nu+1\rangle)\,.
\end{eqnarray}

Another difference to
the Wheeler--DeWitt representation arises for triad operators which in
the Wheeler--DeWitt case would be simply multiplication operators on a
wave function in the metric representation and
thus have continuous spectra. In the loop case, however, the triad
operators
\begin{equation}
 \hat{E}^x = i\frac{\lP^2}{4\pi}\frac{\partial}{\partial K_x}
 \qquad \hat{E}^{\vp} =
 i\frac{\lP^2}{8\pi}\frac{\partial}{\partial K_{\vp}}
\end{equation}
with the Planck length $\lP=\sqrt{8\pi G\hbar}$ have the basis states
(\ref{HomBas}) as eigenstates
\begin{equation}
 \hat{E}^x|\mu,\nu\rangle =
 \case{1}{8\pi}\gamma\lP^2\nu|\mu,\nu\rangle \qquad \hat{E}^{\vp}
 |\mu,\nu\rangle = \case{1}{16\pi}\gamma\lP^2\mu|\mu,\nu\rangle
\end{equation}
and thus discrete spectra (i.e., normalizable eigenstates).
Again, this is different from the Wheeler--DeWitt quantization but
directly analogous to full loop quantum gravity. In particular the
volume $V=4\pi E^{\vp}\sqrt{|E^x|}$ has a quantization with discrete
spectrum with eigenvalues
\begin{equation}
 V_{\mu,\nu}=2\pi (\gamma\lP^2/8\pi)^{3/2}\mu\sqrt{|\nu|}\,.
\end{equation}

\subsection{Inverse triad components}

It is often necessary to quantize inverse powers of densitized triad
components, for instance for matter Hamiltonians or curvature
components. Since the basic triad operators have discrete spectra
containing zero, they do not have densely defined inverses which could
otherwise be used for this purpose. Nevertheless one can proceed, and
in the end have regular properties, by rewriting the classical inverse
in an equivalent way and quantizing the new expression \cite{QSDV}. We
demonstrate this for the spatial curvature given by $^3R=2/|E^x|$ for
which we need an inverse of $E^x$. This can be taken as a measure for
the classical singularity where it diverges. Since there is no direct
way of quantizing this expression via an inverse of $\hat{E}^x$ we
first write
\[
 \frac{E^{\vp}\sgn E^x}{2\sqrt{|E^x|}}= \frac{-1}{8\pi G}\{K_x,V\}=
 \frac{1}{4\pi\gamma G}\tr \tau_3 e^{-\gamma K_x\tau_3}\{e^{\gamma
   K_x\tau_3},V\}
\]
where the first step replaces the inverse power of $E^x$ by only
positive powers occurring in $V$ at the expense of introducing $K_x$
for which we do not have a direct quantization. Nevertheless, in the
next step we obtain an equivalent expression which only contains
exponentials of $K_x$ which we can quantize directly. Using the volume
operator and turning the Poisson bracket into a commutator then yields
a densely defined operator
\begin{eqnarray} \label{ExInv}
 \widehat{\frac{E^{\vp}\sgn E^x}{\sqrt{|E^x|}}} &=&
 \frac{-i}{2\pi\gamma G\hbar} \tr\tau_3 e^{-\gamma K_x\tau_3}[
 e^{\gamma K_x\tau_3},\hat{V}]\\
 &=& \frac{4i}{\gamma\lP^2} (\sin(\case{1}{2}\gamma
 K_x)\hat{V}\cos(\case{1}{2}\gamma K_x)- \cos(\case{1}{2}\gamma
 K_x)\hat{V}\sin(\case{1}{2}\gamma K_x))\nonumber
\end{eqnarray}
with eigenvalues
\begin{equation}
 \frac{2}{\gamma\lP^2}(V_{\mu,\nu+1}-V_{\mu,\nu-1})=
 \case{1}{2}\sqrt{\frac{\gamma\lP^2}{8\pi}} \,\mu
 (\sqrt{|\nu+1|}-\sqrt{|\nu-1|})\,.
\end{equation}
Since $\hat{E}^{\vp}$ has eigenvalues $\gamma\lP^2\mu/16\pi$, we can
write
\begin{equation}
 \widehat{\frac{\sgn E^x}{\sqrt{|E^x|}}}|\mu,\nu\rangle =
 (\gamma\lP^2/8\pi)^{-1/2} (\sqrt{|\nu+1|}-\sqrt{|\nu-1|})
\end{equation}
which has the expected behavior for $|\nu|\gg 1$ but behaves very
differently from the classical expectation for small $|\nu|$.

Taking this as a measure for the singularity indicates that it is
removed in quantum gravity since the eigenvalues remain finite even
when $\nu=0$ at the classical singularity. Nevertheless, a final
confirmation of an absence of singularities can only come from
considerations of the dynamics which must allow us to evolve further
even when we reach a point corresponding to the classical
singularity. Only then can we conclude that the singularity as a
boundary of space-time has been removed.

\subsection{Dynamics}

The spherically symmetric Hamiltonian constraint (\ref{HSphSymm}) can
be used to find the expression for the homogeneous Kantowski--Sachs
interior
\begin{equation} \label{HKS}
 H[N] = (2G)^{-1}N |E^x|^{-1/2}\left(
 (K_{\vp}^2+1) E^{\vp}+2
 K_{\vp}K_x E^x)\right)
\end{equation}
where we used $\Gamma_{\vp}=0$ with homogeneity in
(\ref{Gammap}). There are different terms in this expression, those
quadratic in $K$ and the $K$-independent one which comes from the spin
connection in the curvature of the Ashtekar connection. In the full
theory there would only be curvature components of $A_a^i$ in the
Euclidean part $\epsilon_{ijk}F_{ab}^i E^a_jE^b_k$ of the constraint,
which can be represented via holonomies by using
\[
 s_1^a s_2^bF_{ab}^i(x)\tau_i = (h_{\alpha}-1)/\Delta +O(\Delta)
\]
where $\alpha$ is a small loop of coordinate area $\Delta$ and with
tangent vectors $s_1$ and $s_2$. For small $\Delta$ in a limit
removing a regulator one can use $h_{\alpha}$ as an excellent
approximation for the curvature components, and stick this together
with quantizations of the triad components to obtain a quantization of
the constraint \cite{QSDI}. This is different in a homogeneous context
(or in any symmetric model where some directions are homogeneous)
because we have only exponentials of connection components, but not
holonomies with an adjustable edge length that shrinks in a continuum
limit.  Nevertheless, since the constraint operator in the full theory
is based on holonomies quantizing the $F$-components, this has to be
the case also for symmetric models related to the full theory. The
only possibility to use $h_{\alpha}$ as a good approximation is then
given when the arguments of the exponentials are small in
semiclassical regimes where the classical constraint is to be
reproduced. In other regimes, one does not expect the classical
constraint to be of any value for guidance and in fact usually obtains
strong quantum corrections.

In a semiclassical regime one has small curvature such that the
extrinsic curvature components can be assumed to be small when
checking the classical limit of the constraint. However, Ashtekar
connection components are not necessarily small since for them also the
spin connection plays a role. Here, another difference to the full
theory arises: while in general spin connection components do not have
coordinate independent meaning and in fact can be made arbitrarily
small in any neighborhood, some of the components (such as
(\ref{Gammap}) in the spherically symmetric model) obtain invariant
meaning in a symmetric context where only transformations respecting
the symmetry are allowed. Usually, unless the model has flat symmetry
orbits such that the spin connection vanishes, one cannot expect the
components to be small even in semiclassical regimes. This requires a
special treatment of the spin connection in symmetric models, which is
possible in a general manner \cite{Spin,SphSymmHam}. For this reason
we have started the quantization in this model with extrinsic
curvature components and will also use them instead of Ashtekar
connection components in holonomies when constructing the Hamiltonian
constraint.

One may ask what the relation to the full theory then is where
holonomies of the Ashtekar connection are basic, while holonomies of a
tensor such as extrinsic curvature cannot even be defined. The
arguments presented before explain why extrinsic curvature is
important to analyze the classical limit, but this does not show the
contact to the full theory. This will be much clearer in inhomogeneous
models which are in between homogeneous ones and the full theory.
Here, we will have directions along symmetry orbits, for which the
techniques just described will apply, and inhomogeneous directions for
which we will use holonomies of the Ashtekar connection as in the full
theory. As we will discuss in more detail in the quantization of the
spherically symmetric model, all this fits into a general scheme which
allows to derive expressions in all different classes of models.

We can now express the terms quadratic in curvature components via
holonomies, such as
\[
 K_{\vp}^2+1=-\frac{2}{\gamma^2\delta^2}\tr\tau_3 (e^{-\delta\gamma
   K_{\vp}\tau_1} e^{-\delta\gamma K_{\vp}\tau_2} e^{\delta\gamma
   K_{\vp}\tau_1} e^{\delta\gamma
   K_{\vp}\tau_2}+\gamma^2\delta^2\tau_3)+ O(\delta)
\]
and
\[
 K_xK_{\vp}=\frac{2}{\gamma^2\delta^2} \tr\tau_1 (e^{-\delta\gamma
   K_x\tau_2} e^{-\delta\gamma K_{\vp}\tau_3} e^{\delta\gamma
   K_x\tau_2} e^{\delta\gamma
   K_{\vp}\tau_3})+O(\delta)\,.
\]
Triad components, together with Pauli
matrices in the traces, can be obtained in the right combinations from
the Poisson brackets
\[
 \tau_3\frac{E^{\vp}}{\sqrt{|E^x|}}= -\frac{1}{4\pi\gamma\delta G}
 e^{-\delta\gamma K_x\tau_3}\{e^{\delta\gamma K_x\tau_3},V\}
\]
as already used for (\ref{ExInv}), and
\[
 \tau_1\sqrt{|E^x|}= -\frac{1}{4\pi\gamma\delta G} e^{-\delta\gamma
   K_{\vp}\tau_1}\{e^{\delta\gamma K_{\vp}\tau_1},V\}\,.
\]
In all expressions, besides the volume $V=4\pi\int\md x
\sqrt{|E^x|}E^{\vp}$ only holonomies $h_x^{(\delta)}:=e^{-\gamma\delta
  K_x\tau_3}$, $h_{\vt}^{(\delta)}:=e^{-\gamma\delta K_{\vp}\tau_1}$
and $h_{\vp}^{(\delta)}:=e^{-\gamma\delta K_{\vp}\tau_2}$ of the
symmetric Ashtekar connection (\ref{ASphSymm}), expressed through
extrinsic curvature components, occur which can be quantized directly.

In a more symmetric form, which as we will see later also applies in
general, we write
\begin{eqnarray*}
 &&\frac{(K_{\vp}^2+1)E^{\vp}+2K_xK_{\vp}E^x}{\sqrt{|E^x|}} \sim
 \frac{1}{2\pi\gamma^3\delta^3 G} \tr(
 (h_{\vt}h_{\vp}h_{\vt}^{-1}h_{\vp}^{-1}+ \gamma^2\delta^2\tau_3)
 h_x\{h_x^{-1},V\}\\
 &&\qquad\qquad\qquad+h_xh_{\vt}h_x^{-1}h_{\vt}^{-1} h_{\vp}\{h_{\vp}^{-1},V\}+
 h_{\vp}h_xh_{\vp}^{-1}h_x^{-1} h_{\vt}\{h_{\vt}^{-1},V\})\\
 &&= \frac{1}{4\pi \gamma^3\delta^3 G} \sum_{IJK} \epsilon^{IJK}
 \tr\left((h_Ih_Jh_I^{-1}h_J^{-1}-\gamma^2\delta^2
 F(\Gamma)_{IJ}) h_K\{h_K^{-1},V\}\right)
\end{eqnarray*}
with the curvature components $F(\Gamma)_{IJ}$ of the spin connection,
i.e.\ here
\[
 F(\Gamma)=\md\Gamma=-\tau_3\sin\vt\md\vt\wedge\md\vp
\]
such that only $F(\Gamma)_{\vt\vp}:=i_{X_{\vp}}i_{X_{\vt}}F(\Gamma)=
-\tau_3$ appears, with the symmetry generators
$X_{\vt}=\partial_{\vt}$ and $X_{\vp}=(\sin\vt)^{-1}\partial_{\vp}$.

Quantizing and evaluating the action explicitly through the action of
basic operators leads to a constraint operator of the form
\begin{eqnarray}
 \hat{H}^{(\Le)} &=& -iN
(\gamma^3\Le^3G\lP^2)^{-1}
\sum_{IJK} \epsilon^{IJK}\tr\left(({h}_I^{(\Le)}
{h}_J^{(\Le)}  
{h}_I^{(\Le)-1} {h}_J^{(\Le)-1}-\gamma^2\Le^2F(\Gamma)_{IJ})
{h}_K^{(\Le)} [{h}_K^{(\Le)-1},\hat{V}]\right)\nonumber\\
&=& -2 iN(\gamma^3\Le^3G\lP^2)^{-1}\left(
 8\sin\frac{\Le\gamma K_{\vp}}{2}\cos\frac{\Le\gamma K_{\vp}}{2}
   \sin\frac{\Le\gamma K_x}{2}\cos\frac{\Le\gamma K_x}{2}\right.\nonumber\\
&& \quad\times\left.\left(\sin\frac{\Le\gamma K_{\vp}}{2}\hat{V}\cos\frac{\Le\gamma K_{\vp}}{2}-
 \cos\frac{\Le\gamma K_{\vp}}{2}\hat{V}\sin\frac{\Le\gamma
   K_{\vp}}{2}\right) \right.\nonumber\\
 && + \left.\left(4\sin^2\frac{\Le\gamma K_{\vp}}{2}\cos^2\frac{\Le\gamma
K_{\vp}}{2}+\gamma^2\Le^2\right)\right.\nonumber\\
 && \quad\times\left.
 \left(\sin\frac{\Le\gamma K_x}{2}\hat{V}\cos\frac{\Le\gamma K_x}{2}-
 \cos\frac{\Le\gamma K_x}{2}\hat{V}\sin\frac{\Le\gamma K_x}{2}\right)
\right)
\end{eqnarray}
where $\Le>0$ is regarded as a parameter analogous to the edge length
in the full theory. From the holonomy operators one obtains shifts in
the labels when acting on a state $|\mu,\nu\rangle$ which in the triad
representation given by the coefficients $\psi_{\mu,\nu}$ in a
decomposition $|\psi\rangle =
\sum_{\mu,\nu}\psi_{\mu,\nu}|\mu,\nu\rangle$ leads to the difference
equation
\begin{eqnarray}
 &&0=(\hat{H}^{(\Le)}\psi)_{\mu,\nu}= 2\delta\sqrt{|\nu+2\Le|}
 (\psi_{\mu+2\Le,\nu+2\Le}- \psi_{\mu-2\Le,\nu+2\Le})\label{diffeq}\\
&& +\case{1}{2} (\sqrt{|\nu+\Le|}-\sqrt{|\nu-\Le|})
\left((\mu+4\Le)\psi_{\mu+4\Le,\nu}- 2(1+\gamma^2\Le^2)\mu\psi_{\mu,\nu}+
(\mu-4\Le)\psi_{\mu-4\Le,\nu}\right)\nonumber\\
&&-2\delta\sqrt{|\nu-2\Le|} (\psi_{\mu+2\Le,\nu-2\Le}-
\psi_{\mu-2\Le,\nu-2\Le}) \,.\nonumber
\end{eqnarray}

\begin{figure}
 \includegraphics[width=5cm]{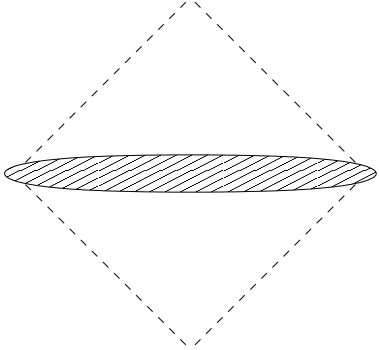} \hfill
\begin{minipage}[b]{8cm}
\caption{Interior of a Schwarzschild black hole with the quantum
  region (hatched) replacing the classical singularity. This allows to
  extend space-time to the new upper region. How these regions are
  embedded in a full space-time is left unspecified here. \label{NonSing}}
\end{minipage}
\end{figure}

We are now in a position to analyze whether or not there is a
singularity in the quantum theory. There are a few key differences to
the usual classical formulation, the first one coming from the fact
that we are using triad variables. Compared to a metric formulation,
this provides us with an additional sign factor $\sgn E^x$ determining
the orientation of space. Accordingly, there are two regions of
minisuperspace separated by the line $E^x=0$ where the classical
singularity would be. We have already seen that the classical
divergence of inverse powers of $E^x$ does not occur in a loop
quantization, but the real test of a singularity can only come from
the dynamics: starting with initial values in one region of
minisuperspace we need to find out whether we can uniquely evolve to
the other side, through the classical singularity. In the quantum
theory this is done for the wave function which we can prescribe for
sufficiently many initial values at some $\nu>0$ and additional
boundary values at $\mu=0$ so as to provide a good initial value
formulation for the difference equation (\ref{diffeq}) as described in
detail in \cite{HomCosmo}. One can then see by direct inspection that
indeed this will uniquely fix the wave function not just at positive
$\nu$ where we started, but also at negative $\nu$, at the other side
of the classical singularity. Thus, quantum geometry automatically
allows us to evolve through the classical singularity which therefore
is removed from quantum gravity.  Intuitively, the region of negative
$\nu$ corresponds to a region of a space-time diagram at the other
side of the singularity, as sketched in Fig.~\ref{NonSing}, which
therefore is no longer a boundary but a region of high curvature where
the classical theory and its smooth space-time picture break down
\cite{BHInt}.

There are many basic aspects which are playing together in just the
right way for this result to hold true. They all come directly from
the loop quantization and are not put in by hand; in fact, they had
been recognized as essential for a background independent quantization
a few years before their role in removing classical singularities
emerged. The loop representation is important in two ways since via
discrete triad spectra it leads to the kinematical results of
non-diverging inverse powers of densitized triad components, and
through the representation of holonomies to the dynamical constraint
as a difference operator. Moreover, the theory is based on densitized
triads which, as discussed before, has consequences for the position
of classical singularities in minisuperspace important for how one can
evolve through them. This automatically provides us with the sign
factor from orientation and thus a region beyond the classical
singularity. Still, also the dynamical law has to be of the right form
for an evolution to this other side of minisuperspace to be possible.

Thus, we have a few essential effects which automatically come from a
loop quantization.  Once recognized and identified, they can easily be
copied in other quantization schemes inspired by loop quantum gravity
and cosmology. However, in such a case one has to guess anew in each
model what the relevant basic properties would be since there is no
underlying scheme for guidance. With the loop quantization we have
such a general scheme which just needs to be evaluated in different
models. Only then can the results be regarded as reliable expectations
for quantum gravity, rather than possibly artificial consequences of
choices made. The sign of triad components, for instance, appears
automatically and then gives rise to the additional side of the
classical singularity to which we can evolve. In loop inspired
approaches without a link to the full theory, however, the sign is
introduced by hand by extending the range of metric variables to
negative values. While this leads to similar results in isotropic
models \cite{BohrADM}, except that the geometrical meaning of the sign
remains unclear, there are differences in the black hole interior
\cite{Modesto}. In particular, this approach would suggest that even
the horizon can be penetrated by the homogeneous quantum evolution
despite the fact that space-time becomes inhomogeneous outside. This
problem does not occur in the quantization described here since the
horizon remains a boundary (corresponding to $\mu=0$).

\subsection{Effective dynamics}

The non-singular quantum dynamics is obviously very different from the
classical one even though they can be shown to agree in classical
regions at large densitized triad components and small curvature
\cite{SemiClass}. In between, there is a regime where equations of
motion of the classical type, i.e.\ ordinary differential equations in
coordinate time, should be able to describe the system even though
quantum effects are already at work. One can think of these equations
as describing the position of wave packets which spread only slightly
in semiclassical regimes \cite{Time,EffAc,Perturb}.  Quantum effects will then
provide modifications, e.g.\ where inverse powers of densitized triad
components occur in a matter Hamiltonian which are replaced by regular
expressions in quantum geometry. This provides different means to
calculate implications of quantum effects which can so far be done in
homogeneous situations.

For instance if we assume the distribution of a matter system
collapsing into a black hole to be isotropic, its outer radius
$Ra(t)$ is
described by a solution $a(t)$ to the Friedmann equation, with $R$
being the coordinate radius where we cut off spatial slices from a
closed FRW model. If we choose a scalar $\phi$ with potential $V(\phi)$
and momentum $p_{\phi}$, we have the Friedmann equation
\begin{equation}
 a(\dot{a}^2+1)= \frac{8\pi G}{3}(\case{1}{2}a^{-3}p_{\phi}^2+
 a^3 V(\phi))
\end{equation}
which develops a singularity corresponding to the part of the final black
hole singularity covered by matter.

The corresponding effective classical equations are modified by
replacing the classically diverging $a^{-3}$ in the matter Hamiltonian
with a regular function $d(a)$ derived from finite inverse scale
factor operators such as (\ref{ExInv}) \cite{Ambig,ICGC}. Including
two ambiguity parameters $j$ (a half integer) and $0<l<1$, this can be
parameterized as
\begin{equation} \label{deff}
  d(a)_{j,l}:= a^{-3} p_l(3a^2/\gamma j\ell_{\rm
   P}^2)^{3/(2-2l)}
\end{equation}
with
\begin{eqnarray}
 p_l(q) &=&\frac{3}{2l}q^{1-l}\left( \frac{1}{l+2}
\left((q+1)^{l+2}-|q-1|^{l+2}\right)\right.\\
 && - \left.\frac{1}{l+1}q
\left((q+1)^{l+1}-{\rm sgn}(q-1)|q-1|^{l+1}\right)\right)\,.\nonumber
\end{eqnarray}
The essential property of $d(a)_{j,l}$ is that it is increasing from
zero for $a^2<\frac{1}{3}\gamma j\lP^2$ which through the Friedmann
equation implies a different dynamical behavior at small volume.  This
model then provides an intuitive explanation for the removal of
classical singularities even at the effective level since the
equations lead to a bounce at non-zero $a$: due to the modified
density the kinetic term is negligible at small $a$, and matter
evolution equations from the modified matter Hamiltonian imply
friction of $\phi$ \cite{Closed}. The potential term is then
dominating and almost constant which means that the bounce is
approximately of de Sitter form \cite{BounceClosed}. In this
interpretation of collapsing matter this means that it does not
collapse completely but rebounds after a point of minimal contraction
is reached.

\begin{figure}
\begin{center}
 \includegraphics[width=5cm]{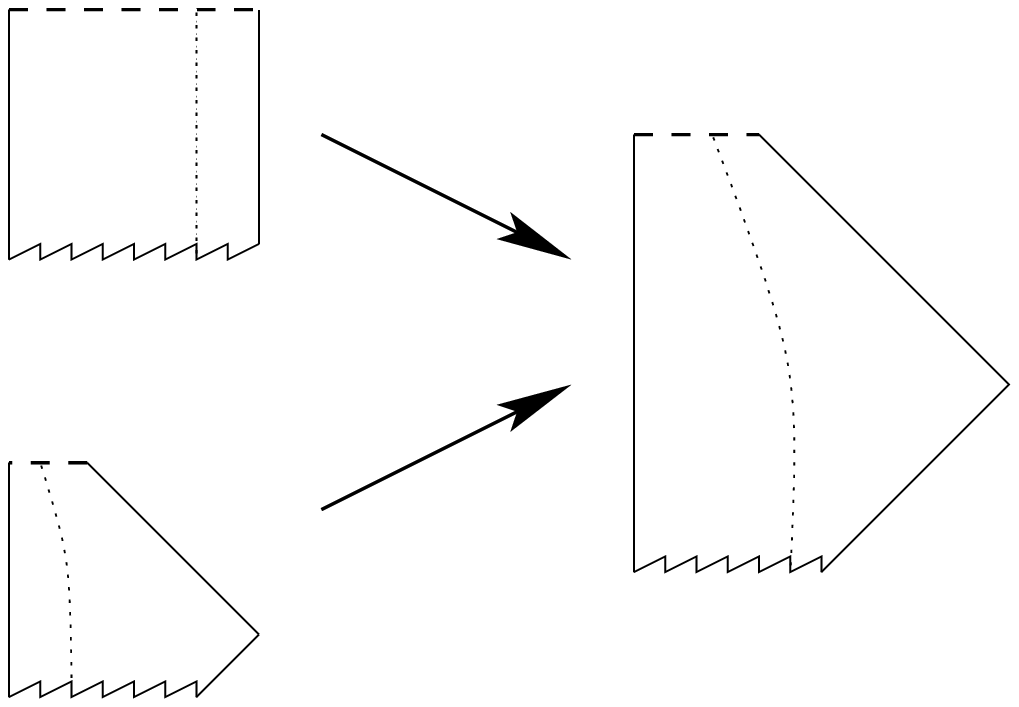}\hspace{2cm}
 \includegraphics[width=5cm]{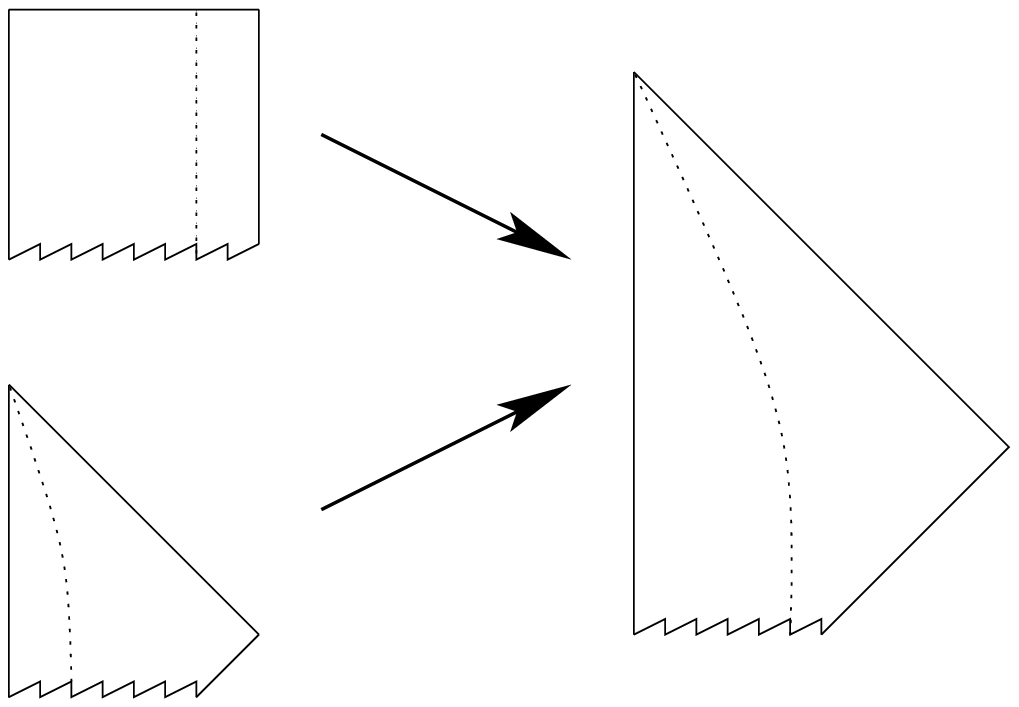}
\end{center}
\caption{A closed Friedmann--Robertson--Walker model and a generalized
  Vaidya metric are matched to form a collapse
  model. Singularities are indicated by dashed lines and the
  matching surface by dotted lines. The bottom parts of the diagrams are
  cut off since they depend on details of the solutions. The left hand
  side shows the classical case with a future singularity, while the
  right hand side shows the singularity-free effective case.
\label{MatchingClass}}
\end{figure}

So far, we had only access to the inside of the matter contribution
which we assumed to be isotropic. The solution can now be matched to a
suitable solution describing the outside, which would be able to tell
us, for instance, whether horizons form. For pressure-less matter one
can match to the static Schwarzschild solution as in the
Oppenheimer--Snyder model \cite{OppSny}. This is the case classically
only for dust, which however can develop pressure if quantum
modifications come into play. (This is in agreement with our earlier
observations that quantum effects will not allow the presence of a
static vacuum solution.) We have thus chosen the more general scalar
matter which has pressure even classically. Physically, pressure leads
to shock waves at the outer boundary giving rise to a non-static
exterior. This can be described by a generalized Vaidya metric
\begin{equation}
 \md s^2=-(1-2M(\chi,v)/\chi)\md v^2+2\md v\md\chi+\chi^2 \md\Omega^2
\end{equation}
which we can match to the interior (Fig.~\ref{MatchingClass}) by
requiring equal induced metric and extrinsic curvature at the time-like
matching surface $\Sigma$ defined by $r=R$ inside and $\chi=\chi(v)$
outside.

In this way we can see what the collapsing matter implies for the
outside space-time at least in a neighborhood \cite{Collapse}. To have
access to the full outside all the way up to an asymptotic observer we
would need to specify the matter content outside. While this is
possible e.g.\ as the same scalar matter as inside, only
inhomogeneous, it would not necessarily be correct physically. In
fact, we have modified only the classical equations describing the
interior, while we did not use effective equations outside. When the
matter distribution extends over a large region in the early stages,
one does not expect strong modifications, but this is not clear close
to the bounce. In this region, the interior equations are strongly
modified, and this is transferred to the outside via the matching
conditions in a rather indirect way: We use the classical generalized
Vaidya metric, but did not specify the matter content. It is
effectively the energy momentum tensor which carries quantum effects
from the interior to the outside via the matching. Prescribing the
outside matter content would remove this transfer and stop us from
seeing possible quantum effects outside.

We write the interior metric as
\[
 \md s^2=-\md t^2+X(r,t)^2\md r^2+Y(r,t)^2\md\Omega^2
\]
with $X(r,t)=a(t)/(1+r^2/4)$ and $Y(r,t)=rX(r,t)$. On the matching
surface $r=R$ of the interior and $\chi=\chi(v)$ of the generalized
Vaidya exterior the metric and extrinsic curvature have to agree. From
the metrics we obtain
\begin{equation} \label{chiR}
 \chi|_{\Sigma}=Y|_{\Sigma}
\end{equation}
and
\begin{equation}\label{dvdt}
 \left.\frac{\md v}{\md t}\right|_{\Sigma} =
\left.(1-2M/\chi-2\md\chi/\md v)^{-1/2}\right|_{\Sigma}
\end{equation}
while the extrinsic curvature, computed again from (\ref{ExtrCurv}), gives us
\begin{equation}\label{c}
\left.\frac{YY'}{X}\right|_{\Sigma} =\left.
\chi\frac{1-2M/\chi-\md\chi/\md v}{\sqrt{1-2M/\chi-2\md\chi/\md
    v}}\right|_{\Sigma} 
\end{equation}
and 
\begin{equation} \label{Ktt}
 0=\partial_vM+\frac{\md^2\chi}{\md v^2}+
 \left(1-\frac{2M}{\chi}-3\frac{\md\chi}{\md v}\right)
 \left(\frac{M}{\chi}-\partial_{\chi}M\right)
\end{equation}
which yields a condition for $\partial M/\partial\chi$
at constant $v$.

With $\md\chi/\md v|_{\Sigma}=\dot{\chi}|_{\Sigma}/\dot{v}|_{\Sigma}$
and (\ref{chiR}) we use (\ref{c}) to write the square root in
(\ref{dvdt}) in terms of $Y'$ and $\dot{Y}$ which leads to
\begin{equation} \label{dvdtR}
 \left.\frac{\md v}{\md t}\right|_{\Sigma}=
\left.\frac{(Y'/X+\dot{Y})}{1-2M/Y}\right|_{\Sigma} \,.
\end{equation}
Using (\ref{chiR}) and defining $c:=Y'/X$, (\ref{c}) becomes
\[
 c^2(1-2M/\chi+2\md\chi/\md v)=(1-2M/\chi-\md\chi/\md v)^2
\]
which with
\[
 (\md\chi/\md v)^2= \dot{Y}^2 (1-2M/\chi-2\md\chi/\md v)
\]
(following from $\md\chi/\md v=\dot{\chi}/\dot{v}$ and (\ref{dvdt}))
gives $c^2=1-2M/\chi+\dot{Y}^2$.  Thus,
\begin{equation}\label{M}
 2M|_{\Sigma}=(Y\dot{Y}^2+Y(1-c^2))|_{\Sigma}\,.
\end{equation}

\begin{figure}
\begin{center}
 \includegraphics[width=6cm]{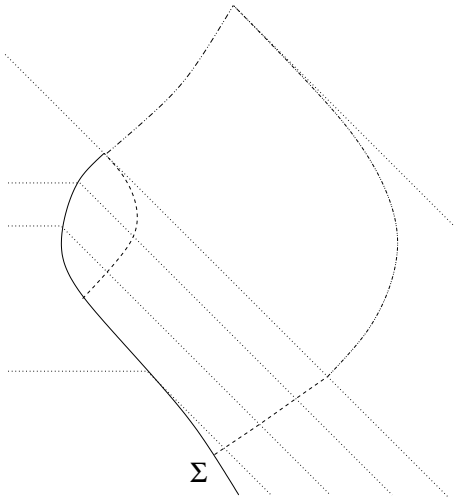}\hspace{5mm}
\begin{minipage}[b]{6cm}
\caption{Sketch of the bouncing effective interior, covered by
  evaporating horizons (dashed). Constant $v$ (outside) and $t$
  (inside) slices are dotted. The matching described in the text only
  refers to the part before the inner horizon collides with the
  expanding matter after the bounce.
\label{Horizon}}
\end{minipage}
\end{center}
\end{figure}

A trapped surface forms in a generalized Vaidya metric when $2M=\chi$,
which lies on the matching surface if $2M=Y$. From (\ref{M}) this
yields the simple condition
\begin{equation}
|\dot{Y}|=c=Y'/X
\end{equation}
which for FRW reduces to
\begin{equation} \label{Hora}
 |\dot{a}|=(1-R^2/4)/R\,.
\end{equation}
Assuming, for now, that this condition will be satisfied at a time
$t(R)$ during collapse, we obtain a horizon covering the bounce
(Fig.~\ref{Horizon}). The squared norm of its normal is given by
$\partial_vM(1-2\partial_{\chi}M)$ which can be computed from
(\ref{Ktt}) using $\md M/\md v=\partial_vM+\partial_{\chi}M\md\chi/\md
v$ and turns out to be zero if the horizon condition (\ref{Hora}) is
satisfied.  The horizon is thus always null when it first intersects
the matching surface.

After the first trapped surface forms on the matching surface,
$|\dot{a}|$ continues to increase before it turns around when the peak
in $d(a)_{j,l}$ is reached. From then on, $|\dot{a}|$ decreases and
reaches $\dot{a}=0$ at the bounce. In between, the trapped surface
condition (\ref{Hora}) will be satisfied a second time at an inner
horizon. Unlike the outer one, it lies in the modified regime where
energy conditions are effectively violated and $\ddot{a}>0$
\cite{Inflation}. It is also null at the matching surface but can
become time-like soon and evaporate later. Similarly, the outer
horizon can become time-like when matter having experienced the
quantum modifications starts to propagate through it. Thus, also the
outer horizon can evaporate and shrink toward the matching surface at
later times, when the inner matter is already expanding.

The horizon thus evaporates and the bouncing matter has a chance to
reappear. Indeed, the condition (\ref{Hora}) will be satisfied also at
a time after the bounce where $\dot{a}$ is now positive.  Thus, the
horizon will intersect again with the matter shells and disappear.  At
such a point, however, the matching breaks down since $\md v/\md t$
diverges when $2M=Y$ and $\dot{Y}>0$. From a single matching of the
interior we obtain only a part of the collapse before a horizon
disappears. At the endpoint of the horizon the interior coordinate $t$
ceases to be good, and we have to match to another patch
(Fig.~\ref{Horizon}).

The precise position of the horizon only follows when we specify the
matter content and field equations (i.e.\ Einstein's equations or
modified ones) and integrate with $M$ and $\partial M/\partial\chi$ as
boundary conditions at the matching surface. Since we leave this open,
we do not get precise information on the horizon but only qualitative
properties. After some time into the collapse, the horizon is expected
to shrink since modifications in the interior imply small violations
of energy conditions (which also allow the bounce to take place).
Radiation of negative energy implies, analogously to Hawking
radiation, that the horizon becomes time-like and shrinks. Later, it
can meet the matching surface again at which point the matter becomes
visible from behind the horizon. If the initial mass was large, it
takes a long time for the bounce to occur and the matter to reemerge
such that for most of the time the system looks like a classical black
hole to an outside observer.

It is not guaranteed that a horizon forms at all since fulfillment of
the horizon condition depends on initial values. In particular,
once we choose $R$ to specify the matching surface in the interior,
Eq. (\ref{Hora}) fixes the value for $\dot{a}$ which needs to be
reached for a horizon to occur. Classically, $\dot{a}$ is unbounded as
we approach the singularity such that the condition will always be
true at one point and there is always a horizon covering the
classical singularity in this model. With the effective equations,
however, $\dot{a}$ is bounded for given initial conditions, and
depending on the value of $R$ it can happen that the horizon condition
is never fulfilled. In this case, there would not be a black hole but
only a matter distribution collapsing to high densities and rebounding.
This rules out black holes of a certain type, in particular those of
small mass: Starting with a configuration such that a horizon forms,
we can decrease $R$ toward zero without changing $a(t)$. The right
hand side of (\ref{Hora}) then increases and at one point the
condition can no longer be fulfilled. Since with decreasing $R$ we
carve out a smaller piece of the homogeneous interior, the total
initial mass is smaller, giving us a lower bound for the mass of
black holes in this model. Precise values have to be derived from more
detailed models, but this argument shows that large, astrophysical
black holes will be unaffected while primordial ones of small masses
do not form \cite{Collapse}.

\section{Extrapolation}

We have seen two results from homogeneous techniques employed in the
preceding section:
\begin{itemize}
 \item At the fully quantum level of the Kantowski--Sachs model
   describing the Schwarzschild black hole interior the singularity is
   absent (Fig.~\ref{NonSing}).
 \item Matter systems allow effective classical equations for
   their collapse such that the classical singularity is replaced by a
   bounce sometimes shrouded by a horizon (Fig.~\ref{Horizon}).
\end{itemize}
Both results have been arrived at with very different techniques, and
have different physical meaning. The first one only applies to the
vacuum case but provides us with a strict result as to how the
classical singularity is replaced in quantum gravity. It directly
shows that general relativity is singular because it relies on the
smooth classical space-time picture. This picture breaks down at high
curvature and has to be replaced by discrete quantum geometry,
providing a non-singular evolution.

The second result works with matter but is more intuitive, only giving
a picture from effective classical equations. It provides a physical,
rather than geometrical explanation for the failure of general
relativity in strong curvature regimes. Singularities in general
relativity can be understood as a consequence of the always attractive
nature of classical gravity: Once matter collapses to a sufficiently
high density, be it an isolated part or the whole universe, there is
nothing to prevent total collapse into a singularity. Viewing the
Friedmann equation, e.g.\ with scalar matter, as describing a
mechanics system with the matter energy density serving as potential
shows this by the fact that the energy density decreases as a function
of $a$ at fixed $\phi$ and $p_{\phi}$, in particular the kinetic term
$\frac{1}{2}a^{-3}p_{\phi}^2$. Thus, there is an attractive force
driving the system toward $a=0$ (or, as usually expressed in
cosmology, positive pressure which thermodynamically is defined as the
negative change of energy with volume). The modification of $a^{-3}$
by the regular function $d(a)$ in (\ref{deff}), which turns around at
a peak value and then approaches zero rather than infinity at $a=0$,
implies that now the energy density increases as a function of volume
at small scales. This can be interpreted as quantum gravity becoming
repulsive at small scales, which can then easily prevent total
collapse into a singularity.  Moreover, at non-zero but small scales
this repulsive component is still active and leads to modified
behavior. For instance, in an expanding universe it implies that the
expansion is accelerated leading directly to inflation
\cite{Inflation}. In the interpretation of collapsing matter, the same
effect makes the horizon shrink after the bounce such that only the
strong quantum region is covered.

\begin{figure}
\begin{center}
 \includegraphics[width=3cm]{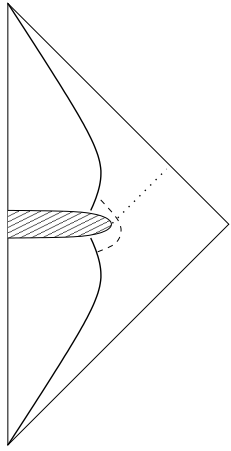} \hspace{5mm}
\begin{minipage}[b]{7cm}
\caption{Combination of Figs.~\ref{NonSing} and \ref{Horizon} where
  the quantum region around the classical singularity is shrouded by an
  evaporating horizon (dashed). The dotted line marks the boundary of
  the part of space-time unaffected by the strong curvature inside.
\label{Combined}}
\end{minipage}
\end{center}
\end{figure}

Since we have used approximations, the question arises how these
partial results can fit into a full picture of quantum black holes.
The first result indicates that space-time can be extended through
classical singularities, but since it gives us access only to the
interior, it is not clear if the new region we reach can also be
accessed from an outside observer. (If not, the black hole would
appear as a wormhole through which one can travel into a new region of
the universe.) The second result now indicates that we can in fact
access the new region since there is only one matching region outside
the collapsing matter, suggesting a picture as in
Fig.~\ref{Combined}. Quite similar ideas have been put forward, by
different motivations, in
\cite{ClosedHorHighDer,ClosedHor,ClosedHorTrapping}.

However, here it is important to bear in mind that we only effectively
described matter outside falling into the collapsing shells. In a more
realistic model, there would be such inhomogeneous matter colliding
with the homogeneous core and making it more heavy. This can then lead
to singularities forming in the outside region. For their resolution
we would again have to use quantum geometry and face the same problem
as to whether or not this will lead to a new region accessible from
the outside.

It is clear that a decisive answer can only be obtained with
inhomogeneous techniques, which we are going to describe in the next
section. Still, even at this level one can see that there are only a
few possible scenarios which can be distinguished by using
inhomogeneous properties of quantum geometry. Irrespective of which
outcome inhomogeneous models will show us, one can already see special
features of quantum geometry leading to a new paradigm about black
hole evaporation. For the first time, this takes into account a
resolution of the classical singularity with implications for apparent
loss of information \cite{BHPara}. In fact, while Hawking radiation
still emerges in a neighborhood of the dynamical horizon and is still
approximately thermal, this is by no means everything coming out of
the black hole at late times. Infalling matter now evolves through the
quantum region of high curvature and reappears later, restoring
correlations which are not recovered by Hawking radiation alone. In
particular, there is no reason for the final state measured on all of
future null infinity to be mixed if we started with a pure initial
state. In the usual picture one would cut out the quantum region (or
the place of the classical singularity) and consider the
future space-time without allowing penetration through that region.
Future null infinity then stops at the intersection with the dotted
line in Fig.~\ref{Combined}, and a state retrieved at this part of
null infinity is indeed mixed since it is obtained by averaging over
the rest to the future. In this way, both the singularity problem and
the information loss paradox are resolved by loop quantum gravity.

\section{Inhomogeneous techniques}

For the spherically symmetric model we need to perform the loop
quantization for inhomogeneous configurations (\ref{K}) and (\ref{E})
such that the basic fields now depend on the radial coordinate $x$.
Instead of using states such as (\ref{HomBas}) with a finite number of
labels, we now have a field theory with infinitely many kinematical
degrees of freedom. An orthonormal basis of states is given by
\cite{SphSymm}
\begin{equation} \label{SphSymmBas}
 \langle A_x,K_{\vp}|\ldots ,
 k_n,\mu_n,k_{n+1},\mu_{n+1},\ldots\rangle = \prod_n
 \exp(\case{1}{2}ik_n\smallint_{e_n} A_x\md x) \exp(-i\mu_n\gamma
 K_{\vp}(v_n))
\end{equation}
with countably many labels $k_n\in\Z$ and $0\leq \mu_n\in\R$ labeling
edges $e_n$ and vertices $v_n$, respectively, of a 1-dimensional graph
in the radial line. Note that, as already indicated before, we are
using exponentials of the extrinsic curvature component $K_{\vp}$
along homogeneous directions but holonomies of the connection
component $A_x$ along the inhomogeneous direction. Both exponentials
are represented as multiplication operators.

Spatial geometry is encoded in densitized triad operators acting by
\begin{eqnarray}
 \hat{E}^x(x)|\ldots ,
 k_n,\mu_n,\ldots\rangle  &=& \frac{\gamma\lP^2}{8\pi}
\frac{k_{n^+(x)}+k_{n^-(x)}}{2}|\ldots ,
 k_n,\mu_n,\ldots\rangle  \label{Exspec}\\
 \int_{\cal I}\hat{E}^{\vp}|\ldots ,
 k_n,\mu_n,\ldots\rangle &=& \frac{\gamma\lP^2}{8\pi}
\sum_{v_n\in{\cal I}} \mu_n |\ldots ,
 k_n,\mu_n,\ldots\rangle\label{Epspec}
\end{eqnarray}
where $n^{\pm}(x)$ is the edge label to the right (left) of $x$, and
${\cal I}$ is an interval on the radial line (over which we need to
integrate $E^{\vp}$ since it is a density). As before in the
homogeneous case we also obtain densely defined operators for inverse
powers of the triad components, which can in particular be done for
the inverse of $E^{\vp}$ in the spin connection component (\ref{Gammap}).

\subsection{Hamiltonian constraint}

For the Hamiltonian constraint (\ref{HSphSymm}) we again have to obtain
curvature components from holonomies, now using holonomies of $A_x$
for the inhomogeneous radial direction and exponentials of $K_{\vp}$
for homogeneous directions along symmetry orbits. Terms containing the
spin connection component $\Gamma_{\vp}$ belonging to homogeneous
directions will be quantized separately. One may wonder if this
procedure will easily give the right components in the Hamiltonian
constraint, given that it has the rather simple expression
(\ref{HSphSymm}) in terms of extrinsic curvature components while we
are using the Ashtekar connection component $A_x$. As we will see, the
general scheme will yield automatically the right combination of
components by a straightforward construction of loops to be used in
holonomies.

To see this in detail, we first note the difference between $A_x$ and
$K_x$, which is given by the $x$-component of the spin connection. For
a general spherically symmetric triad it takes the form
\begin{equation}\label{Gamma}
 \Gamma = -\eta'\tau^3\md x+
 \frac{E^{x\prime}}{2E^{\vp}}\Lambda \md\vt-
 \frac{E^{x\prime}}{2E^{\vp}}\bar\Lambda \sin\vt\md\vp+ \tau_3\cos\vt\md\vp
\end{equation}
as in (\ref{SphSymmSpin}) with (\ref{GammaComp}).  Here, we recognize
(\ref{Gammap}) as used before as the component along homogeneous
directions. This component is a scalar (noting that both $E^{x\prime}$
and $E^{\vp}$ are densities of weight one), while the $x$-component
$\Gamma_x=-\eta'$ does not have covariant meaning and indeed can be
made arbitrarily small locally by a suitable gauge transformation.
This is analogous to the situation in the full theory, while the gauge
invariant meaning of $\Gamma_{\vp}$ mimics homogeneous models.

Even though $\Gamma_x$ can be made arbitrarily small locally by a
gauge transformation, we cannot assume this when constructing a
suitable Hamiltonian constraint operator. Thus, it must be built into
the construction so as to combine with $A_x$ from radial holonomies to
give $K_x$ as in the expression for the constraint. This
$K_{\vp}K_x$-term in the constraint can, according to the general
construction where connection or extrinsic curvature components derive
from closed holonomies, only come from a loop which has one edge along
a symmetry orbit and one in the radial direction. Starting in a point
$v$, such a holonomy is of the form
$h_x^{(\Le_x)}h_{\vp}^{(\Le)}(v_+)(h_x^{(\Le_x)})^{-1}
(h_{\vp}^{(\Le)}(v))^{-1}$ with a new vertex $v_+$ displaced from $v$
by a coordinate distance $\Le_x$ of the radial edge. (We distinguish
between $\Le_x$ for the radial direction and $\Le$ for the angular
directions since the continuum limit is technically different in both
cases.) This term appears together with $\bar\Lambda(v)$ (coming from
quantizing the triad components) in a trace whose expansion in $\Le$
\begin{eqnarray} \label{hxhp}
 &&-2\tr(h_xh_{\vp}(v_+)h_x^{-1}h_{\vp}(v)^{-1} \bar
 \Lambda(v))\nonumber\\
 &&\quad\sim  -2\gamma\delta
 (K_{\vp}(v_+)\tr(\Lambda(v_+)\bar\Lambda(v))
 + 2K_{\vp}(v_+) \smallint A_x\md x
 \tr(\tau_3\Lambda(v_+)\bar\Lambda(v))) \nonumber\\
 &&\quad= \gamma\delta (K_{\vp}(v_+)\sin(\eta_+-\eta)
 + K_{\vp}(v_+)\smallint A_x\md x\cos(\eta_+-\eta))\nonumber\\
 &&\quad= \Le\Le_x
 \gamma K_{\vp}(v)(A_x(v)+\eta'(v))+O(\Le^2)
\end{eqnarray}
has all the right terms, with $A_x$ coming directly from the radial
holonomy and $\eta(v_+)$ in $\Le\eta'\sim \eta(v_+)-\eta(v)$ from the
internal direction $\Lambda(v_+)$ at the new vertex.  The other term
of the form $K_{\vp}^2$ is obtained from angular holonomies only, as
in the homogeneous case,
\begin{equation} \label{hthp}
 -2\tr(h_{\vt}h_{\vp}h_{\vt}^{-1}h_{\vp}^{-1} \tau_3) \sim
 \Le^2\gamma^2K_{\vp}^2\,.
\end{equation}
The matrices $\tau_3$ and $\bar\Lambda(v)$ in the traces come
from Poisson brackets expressing triad components as in the
homogeneous constraint. Moreover, the spin connection components in
(\ref{HSphSymm}) are again expressed through the curvature
\[
 F(\Gamma) = -\Gamma_{\vp}'\Lambda\md x\wedge\md\vt+
 \Gamma_{\vp}'\bar\Lambda\sin\vt\md x\wedge\md\vp+
 (\Gamma_{\vp}^2-1)\tau_3\sin\vt\md\vt\wedge\md\vp
\]
of the spin connection. Through
\[
\sum_{IJK}\epsilon^{IJK}\tr(\delta_I\delta_JF(\Gamma)_{IJ}
h_K\{h_K^{-1},V\})\propto (\Gamma_{\vp}^2-1)\{A_x,V\}-
2\Gamma_{\vp}'\{-\gamma K_{\vp},V\})
\]
we obtain the additional terms of the constraint, where we included
the length parameters $\delta_I$ (i.e.\ $\delta_x$ for $I=x$ and
$\delta$ for $\vt$ or $\vp$).

This demonstrates how the general procedure works without additional
input: We use exponentials of extrinsic curvature components for
homogeneous directions and holonomies of Ashtekar connection
components for inhomogeneous ones as dictated by the background
independent representation. Spin connection components for
inhomogeneous directions then come in the right form to combine with
extrinsic curvature components, while those in homogeneous directions
are split off and quantized separately. This is possible because those
components, in contrast to the inhomogeneous ones, do have covariant
meaning. This ties together the constructions in homogeneous models
and the full theory, and at the same time opens a direct route to
effective classical equations: Homogeneous spin connection components
usually contain inverse powers of the densitized triad, such as
(\ref{Gammap}). When they are quantized, the classical divergence will
be removed implying modifications at small scales. in homogeneous
models this has been used, e.g., in the Bianchi IX case where it has
been shown to remove the classical chaos
\cite{NonChaos,ChaosLQC}. Similarly, one can use this mechanism to
derive effective classical equations for the spherically symmetric
model and find possible consequences.

Before trusting those effective equations in the inhomogeneous case
one needs to make sure that there is a well-defined Hamiltonian
constraint operator emerging from the procedure described here. So
far, we have only discussed those holonomies and spin connection
components which give us the contributions to the constraint, but they
must now be stuck together with quantizations of triad components so
as to build a well-defined operator for the whole expression.
Moreover, since the expression (\ref{HSphSymm}) is an integrated
density, one has to discretize the integration first and then, after
quantizing the individual terms, perform the continuum limit removing
the regulator. The discretization had already been understood above,
with $v$ and $v_+$ being the endpoints of a discrete interval of size
$\Le_x$, and we convinced ourselves that the continuum limit of the
discretization will yield the correct result. In more detail, one
writes
\[
 H[N]=\int\md xN(x) {\cal H}(x)\sim \sum_n\delta_x^{(n)}N(v_n){\cal
   H}(v_n)
\]
where we discretized the radial line into intervals of coordinate
length $\delta_x^{(n)}$, each one containing the point $v_n$.
Classically, both expressions can be made to agree for any subdivision
by choosing points $v_n$ in the intervals according to the mid point
theorem. Alternatively, if one wants to fix the $v_n$ to be endpoints
of the intervals, the discretization agrees with the classical
constraint in the continuum limit in which $n\to\infty$ and
$\delta_x^{(n)}\to0$ for all $n$.

For the Hamiltonian constraint in general each term in the sum then
has contributions of the form
\begin{equation} \label{HCont}
 \delta_x {\cal H}(v)\propto \sum_{I,J,K} \epsilon^{IJK}\tr(h_{IJ}
 h_K[h_K^{-1},\hat{V}])
\end{equation}
where we sum over triples $(I,J,K)$ of independent directions which in
symmetric models are given by generators of the symmetry
transformations ($(\vt,\vp)$ in the spherically symmetric case), and
in the full theory or inhomogeneous directions of a symmetric model by
edges of a graph ($x$ in spherical symmetry). The holonomies $h_{IJ}$
are formed according to the symmetry: if both directions $I$ and $J$
are inhomogeneous, $h_{IJ}$ is a holonomy along a closed loop
$\alpha_{IJ}$ constructed from edges in the $IJ$-plane of a graph to
act on; if at least one of the two directions is homogeneous,
$h_{IJ}=h_Ih_Jh_I^{-1}h_J^{-1}- \gamma^2\delta_I\delta_J \hat{F}_{IJ}$
with $h_I$ being exponentials of the (su(2)-valued) extrinsic
curvature components belonging to the $I$-direction for a homogeneous
direction $I$ or holonomies along an inhomogeneous direction. (The
appearance of $F(\Gamma)$ can be understood as a correction term since
loops made from holonomies along vector fields generating symmetries
do not close if orbits have non-zero curvature \cite{cosmoIII}.) The
size of loops $\alpha_{IJ}$ or single holonomies is determined by the
size $\delta_x$ of the discretization.  The final holonomy $h_K$
either belongs to an edge transversal to both directions $I$ and $J$,
or again to exponentiated extrinsic curvature components if $K$ is
homogeneous.  These combinations are chosen in such a way that
$h_{IJ}$ yields the correct curvature components, the commutator gives
the necessary triad components, and both terms together provide just
the right product of lattice sizes such as $\Le_x$ in order for the
sum to take the form of a Riemann summation of the original integral.
In the full theory, this procedure only results in the so-called
Euclidean part of the constraint which can be used to construct the
Lorentzian constraint \cite{QSDI}. In the models used here, however,
the prescription (\ref{HCont}) is sufficient even for Lorentzian
signature.

This can now be illustrated and applied in the spherically symmetric
model where the first case above cannot appear since there is only one
inhomogeneous direction. We have thus two cases, one in
which direction $I$ or $J$ is radial, resulting in the first product
of holonomies discussed above, combined with a commutator
\begin{eqnarray*}
 h_{\vp}[h_{\vp}^{-1},\hat{V}] &=& \hat{V}-\cos\case{1}{2}\gamma K_{\vp}
 \hat{V}\cos\case{1}{2}\gamma K_{\vp}- \sin\case{1}{2}\gamma K_{\vp}
 \hat{V}\sin\case{1}{2}\gamma K_{\vp}\\
 &&- 2\Lambda (\cos\case{1}{2}\gamma K_{\vp}
 \hat{V}\sin\case{1}{2}\gamma K_{\vp}-\sin\case{1}{2}\gamma K_{\vp}
 \hat{V}\cos\case{1}{2}\gamma K_{\vp})\,.
\end{eqnarray*}
In the second case we have the other product of holonomies and a
commutator
\begin{eqnarray*}
 h_x[h_x^{-1},\hat{V}] &=& \hat{V}-\cos\case{1}{2}\smallint A_x
 \hat{V}\cos\case{1}{2}\smallint A_x- \sin\case{1}{2}\smallint A_x 
 \hat{V}\sin\case{1}{2}\smallint A_x\\
 &&+ 2\tau_3 (\cos\case{1}{2}\smallint A_x
 \hat{V}\sin\case{1}{2}\smallint A_x-\sin\case{1}{2}\smallint A_x 
 \hat{V}\cos\case{1}{2}\smallint A_x)\,.
\end{eqnarray*}
The integration here is over an interval of size $\Le_x$ such that in
the limit of a fine discretization this term is of order $\Le_x$ as
needed for the Riemann sum.

What we did not specify yet is how the discretization is adapted to a
graph the constructed operator is supposed to act on, i.e.\ whether
$v$ and maybe $v_{\pm}$ are already vertices of the graph or arbitrary
points. At this point, choices need to be made which lead to different
versions of the constraint. The same choices arise in the full theory
\cite{LM:VertSm,Consist,Master}, but they can be studied much more easily
in the spherically symmetric model such that it may be possible to
rule out some versions.

It is already non-trivial to check that different versions lead to
well-defined operators at all. For this, the action after performing
the continuum limit, in which the number of discretization points
becomes infinite, must be finite. If the action were non-zero at each
discretization point, there would not be a well-defined operator in
the limit and the regulator could not be removed. One would then only
deal with a lattice regulated theory rather than a quantization of the
continuum theory. In the full theory, there is a well-defined operator
because the action of the constraint is zero unless a discretization
point is already a vertex. Starting with states with finitely many
vertices then leads to a densely defined operator. This comes about in
the full theory because the constraint contains the volume operator in
such a way that it acts only on planar vertices if there is no vertex
already present in the graph. Since the full volume operator
annihilates all planar vertices, there are only finitely many
contributions from the vertices already present.

In spherical symmetry, however, all vertices are planar since graphs
are just 1-dimensional. This simple general argument is thus not
available and it is not obvious that the same construction scheme will
result in a well-defined operator. It turns out, however, that this is
the case as a consequence of how triad components in the constraint
are quantized: a discretization point which is not already a vertex of
the graph to act on will be annihilated such that only finitely many
contributions from the vertices remain. One can thus use the same type
of operator, just with adaptations to the symmetric
situation. 

Nevertheless, one can also choose different constructions where the
discretization is given directly by the graph, i.e.\ discretization
intervals would be complete edges of the graph. Since endpoints of
discretization intervals are always vertices of a state, the continuum
limit would then require also states to change and become finer and
finer.  In this picture, the continuum limit of the constraint
operator can only be tested on states which are suited to the
continuum behavior, while there are also other states where
discreteness is essential and where the classical constraint would be
corrected from quantum effects. Moreover, in the continuum limit the
number of vertices diverges and the constraint operator becomes
ill-defined just as the usual Wheeler--DeWitt operator is. Both
schemes result in well-defined operators, but they lead to quite
different equations of motion and require different conceptual
viewpoints about the continuum limit.  When the continuum limit is to
be ensured for each state, one requires in a sense that the classical
equations are sensible at arbitrarily small scales, and corrections
could only come from quantum uncertainties. In the second picture, on
the other hand, the classical continuum picture arises only after a
certain coarse graining, or by working only with states which are not
sensitive to the microscopic structure. If one chooses a state which is
sensitive to small scales, or looks very closely at small scales of
even a semiclassical state, then corrections to the classical
expressions arise, for instance as a consequence of the underlying
discreteness.  This second viewpoint has been taken successfully in
cosmological models, and is, as we will see, also fruitful in black
hole models.

\subsection{Dynamics}

Since the constraint operator is again constructed from holonomies
which act by shifting the labels, it implies difference equations for
states in a triad representation. (Note that also in the spherically
symmetric model the triad representation exists, unlike in the full
theory where flux operators do not commute \cite{NonCommFlux}.) These
equations are now not only partial difference equations but also have
many independent variables. Interestingly, the type of difference
equation is very different for the two versions of the constraint
operator: in the second case the number of edges and vertices of the
original state is unchanged and the operator only acts on the labels.
This results in difference equations with independent variables
$k_e$ for each edge and $\mu_v$ for each vertex. Since the operator
does not change the number of edges and vertices, one obtains coupled
difference equations in a fixed number of variables for each sector
given by the number of vertices.

In the first version of the constraint, however, the situation is very
different. Now, new vertices are created and edges split in each
action of the operator. Thus, for a triad representation it is not
enough to work with a fixed number of vertices. Rather, all graphs
have to be taken into account for the equation, which implies that one
has to deal with infinitely many independent variables and thus
functional difference equations. 

We thus return to the simpler type of difference equation implied by
the other version of the constraint and discuss what one can already
say about the singularity issue. First of all, one will have to
identify classical singularities on minisuperspace in order to study
the constraint equation in a neighborhood. In an isotropic model this
is simple since the only way is for the volume to go to zero
\cite{HomCosmo}.  Similarly, one can identify the classical
singularity on an anisotropic minisuperspace where all densitized
triad components would go to zero. The situation is not so clear in
midisuperspaces such as spherical symmetry since there are more
possibilities for a singularity to develop. Even though this has not
been settled in general, there are many cases where a singularity is
characterized by $E^x$ approaching zero (which on classical solutions
such as Schwarzschild implies that $E^{\vp}$ becomes zero, too). This
is also in agreement with the general mechanism removing singularities
seen so far in homogeneous models: there it is the sign coming from
orientation which leads to different regions of minisuperspace
separated by the classical singularity. The quantum evolution, as we
have seen in the Kantowski--Sachs model, can then allow us to evolve
between the two regions, thus removing the classical singularity as a
boundary.

The role of orientation is now played by $\sgn E^x$ since $\det q=E^x
(E^{\vp})^2$. Since $E^x$ depends on the radial position, or the edge
after quantization, the boundary of our midisuperspace has many
components which we can identify with an inhomogeneous classical
singularity, corresponding to the fact that inhomogeneous
singularities behave differently in their different points. For the
states this means that we encounter a section of a classical
singularity each time an edge label $k_e$ becomes zero. As in
homogeneous models, we can then use the evolution equation in the
triad representation in order to see if an evolution through this part
of the boundary is possible. Since the structure of the difference
equation for a given edge label is very similar to the homogeneous
equation, one can expect that the boundary indeed disappears and that
the quantum evolution connects regions of midisuperspace corresponding
to different local orientations. There would thus be no singular
boundary, and the same mechanism as in homogeneous models could also
remove spherically symmetric singularities.

This scenario has been verified in \cite{SphSymmSing}, noting a
crucial difference to homogeneous models wich require a symmetric
ordering of the constraint. Thus, quantization choices are reduced by
looking at less symmetric models, so far in such a way which maintains
the validity of the general picture. Yet, there are also open issues
left for a general understanding. For instance, while the results are
independent of the matter Hamiltonian and can be extended to
cylindrical gravitational wave models, thus also allowing local
degrees of freedom, they are so far based only one type of the
constraint which leads to difference equations easier to deal with.
The behavior with the other version is not easy to see, but if it is
singularity free, too, the mechanism is likely to be different. Most
importantly, the kind of initial/boundary value problem suitable for
the constraint equations needs to be analyzed in more detail to
guarantee that there are suitable and sufficiently many solutions with
the correct classical limit. At this point the anomaly issue, i.e.\
whether two constraint operators with different lapse functions have
the correct commutator, becomes important. These considerations thus
provide a promising and treatable way to distinguish different
versions of the quantization by their physical implications, which can
then be extrapolated to the full theory.
\subsection{Horizons}

A feature of black holes which is new compared to cosmological models,
and which requires inhomogeneous situations, is given by the presence
of horizons. Global concepts such as the event horizon are, of course,
not helpful in our case since we would need to solve the Hamiltonian
constraint completely before being able to discuss this issue. There
are more practical definitions such as apparent horizons which,
however, are much more general and do not distinguish between fully
dynamical situations and almost static systems. The quantum behavior
would be most easy to analyze if we can define horizons locally and in
a controlled manner which does not require the full dynamics at once.
Such a concept is presented by isolated \cite{HorRev} or slowly
evolving dynamical horizons \cite{SlowHor}, which even quite
unexpectedly simplify the spherically symmetric Hamiltonian constraint
in their neighborhood.

\subsubsection{Definition}

There are three main parts to the definition of an isolated horizon
$\Delta$ with spatial sections $S\cong S^2$ of given area $a_0$,
embedded into the space manifold $\Sigma$ by $\iota\colon S\to\Sigma$
\cite{IHPhase,ALRev}:
\begin{enumerate}
\item[(i)] The canonical fields $(A_a^i,E^a_i)$ on the horizon are
  completely described by a single field $W=\frac{1}{2}\iota^*A^ir_i$
  on $S$ which is a U(1)-connection obtained from the pull-back of the
  Ashtekar connection to $S$. Here, $r_i$ is an internal direction on
  the horizon chosen such that $W$ is a connection in the spin bundle
  on $S^2$ and $r^iE_i^a = \sqrt{\det q}\, r^a$ on the horizon with
  the internal metric $q$ on $S$ and the outward normal $r^a$ to $S$
  in $\Sigma$.
\item[(ii)] The intrinsic horizon geometry, given by the pull-back of the
 2-form $\Sigma^i_{ab}:=\epsilon_{abc}E^c_i$ to $S$, is determined by
 the curvature $F=\md W$ of $W$ by
\begin{equation} \label{IsoCond}
 F=-\frac{2\pi}{a_0}\iota^*\Sigma^i r_i\,.
\end{equation}
\item[(iii)] The constraints hold on $S$.
\end{enumerate}

A further consequence of the isolated horizon conditions
\cite{IHPhase} is that the curvature ${\cal F}$ of the pull-back of
$A_a^i$ to $S$ has to equal the curvature of $W$: $r_i{\cal
  F}(\iota^*A^i)=2\md W$. This can be seen as one of the
distinguishing features of an isolated horizon since even slowly
evolving horizons at rate $\epsilon$ (related to the expansion of
horizon cross sections \cite{SlowHor}) will break it, though just by
an amount of the order $\epsilon^2$.

When the horizon is introduced as a boundary, condition (i) is used to
identify the horizon degrees of freedom represented by the field
$W$. Condition (ii) then shows that these degrees of freedom are fields
of a Chern--Simons theory on the horizon. It is the main condition
since it relates the horizon degrees of freedom to the bulk geometry,
which after quantization selects the relevant quantum states to be
counted. Condition (iii), on the other hand, does not play a big role
since an isolated horizon as boundary implies a vanishing lapse
function on $S$ for the Hamiltonian constraint which then is to be
imposed only in the bulk.

Thus, when computing black hole entropy in this way, as we will
describe later, the Hamiltonian constraint does not play any role
since it does not act at the boundary, and the Hamiltonian generating
evolution along the horizon need not be considered. In the spherically
symmetric model one can hope that the constraint is simple enough for
an application in this case, either to generate evolution or to impose
the horizon not as a boundary but inside space such that the
constraint would have to be imposed.  In the latter case, moreover, we
will not be able to have an independent boundary theory which is then
matched to the bulk, but would have to find the relevant degrees of
freedom within the original quantum theory.

\subsubsection{Spherical symmetry}

We can now evaluate the conditions for spherically symmetric
connections of the form
\begin{eqnarray}
 A &=& A_x(x)\tau_3\md x + A_{\vp}(x) \bar\Lambda^A(x) \md\vt +
 A_{\vp}(x)\Lambda^A(x) \sin\vt\md\vp + \tau_3\cos\vt\md\vp 
\end{eqnarray}
and densitized triads (\ref{E}), where in general the internal
directions $\Lambda^A$ and $\Lambda$ are different. The connection
component $A_x$ has been discussed before, while the relation
$A_{\vp}\Lambda^A = \Gamma_{\vp}\bar\Lambda-\gamma K_{\vp}\Lambda$,
following from the definition of the Ashtekar connection together with
(\ref{SphSymmSpin}), (\ref{GammaComp}) and (\ref{K}), implies
$A_{\vp}^2=\Gamma_{\vp}^2+\gamma^2 K_{\vp}^2$.

We choose $r_i:=\sgn(E^x)\delta_{i,3}$ such that in fact $r^iE^a_i =
|E^x|\sin\vt\partial_x$ with the intrinsic horizon area element
$|E^x|\sin\vt$ of a metric $|E^x|\md\Omega^2$. Thus,
$W=\frac{1}{2}r_i\iota^*A^i = \frac{1}{2}\sgn(E^x) \cos\vt\md\vp$
whose integrated curvature given by $\oint_S\md W = -2\pi\sgn(E^x(x))$
agrees with the Chern number of the spin bundle, depending on the
orientation given by $\sgn(E^x)$.

Evaluating (\ref{IsoCond}) first shows that in the spherically
symmetric context it is not restrictive since we have $a_0=4\pi
|E^x(S)|$ and the right hand side given by $-\frac{1}{2}\sgn(E^x(S))$
equals $F$ for all $E$. This is not surprising since the spherically
symmetric intrinsic geometry of $S$ is already given by the total area
which is fixed from the outset. (What is free is the sign of $E^x(S)$,
or orientation, which confirms ideas of \cite{Orientation}.) Now the
first condition plays a major role, which we evaluate in the form
$r_i{\cal F}(\iota^*A^i)=2\md W$ \cite{IHPhase}. Since ${\cal
  F}(\iota^*A)= (A_{\vp}^2-1)\tau_3\sin\vt\md\vt\wedge\md\vp$, the
condition requires $A_{\vp}=0$ which will be the main restriction we
have to impose on quantum states in addition to the constraints. This
condition $A_{\vp}=0$ selects 2-spheres in a spherically symmetric
space-time corresponding to cross-sections of a horizon. Indeed, for
the Schwarzschild solution we have $A_{\vp}=\Gamma_{\vp}$ since the
extrinsic curvature vanishes. With (\ref{Gammap}) and the
Schwarzschild triad we obtain the correct condition $x=2M$ for the
horizon. In general,
$A_{\vp}^2=\Gamma_{\vp}^2+\gamma^2K_{\vp}^2=0$ implies
$\Gamma_{\vp}=0$ and $K_{\vp}=0$.

A slowly evolving horizon at rate $\epsilon$ satisfies the condition
$r_i{\cal F}(\iota^*A^i)=2\md W$ only up to terms of the order
$\epsilon^2$. Thus, $A_{\vp}$ is not exactly zero but must be small of
order $\epsilon$, which then is true also for $\Gamma_{\vp}$ and
$K_{\vp}$.

\subsubsection{Dynamics}

In spherical symmetry we can locate a horizon on a state
\cite{Horizon}, which must be at a vertex in order for $K_{\vp}=0$ to
be sharp enough. The condition that $K_{\vp}$ be zero for an isolated
horizon or small for a slowly evolving horizon then leads to important
simplifications which allow a perturbative treatment of the dynamics
around the horizon. Indeed, when acting with the constraint at the
horizon both terms made from holonomies $h_{IJ}$ contain factors of
$\sin\frac{1}{2} \gamma K_{\vp}$ at the horizon vertex or a
neighboring one which must then be small. Ignoring those terms in an
approximation leads to an operator which is diagonal on the spin
network states (\ref{SphSymmBas}) and thus easy to solve as a
constraint or to use for generating time evolution at a boundary. The
additional terms ignored in this approximation can then be included in
a perturbative treatment of the near horizon dynamics.

Without many calculations this already shows how the horizon
fluctuates dynamically. Classically an isolated horizon has constant
area which thus commutes with the Hamiltonian constraint. This is also
true at the quantum level to leading order of the above
approximation since the area operator $\hat{A}(S)=4\pi|\hat{E}^x(S)|$
has the same eigenstates (\ref{SphSymmBas}) as the leading order
constraint. Thus, at this level the horizon area is an observable not
just when the horizon is treated as a boundary, but also if its full
neighborhood is quantized. However, there are additional terms which
arise in higher orders of the perturbation scheme. There are two
reasons for horizon area fluctuations even in the isolated case: While
classically $K_{\vp}=0$ exactly at the horizon and only this value is
important, the quantization does not allow this to hold arbitrarily
sharply. Otherwise, the volume of a shell around the horizon, which
depends on the conjugate momentum $E^{\vp}$ of $K_{\vp}$ could not be
sharp independently of the mass which would contradict semiclassical
properties to hold true at least for massive black holes. Secondly,
the constraint operator acting at the horizon itself depends on
neighboring values of $K_{\vp}$ through $h_{\vp}(v_+)$ in
(\ref{hxhp}). This would give non-zero contributions even if $K_{\vp}$
at the horizon would be zero exactly.

Both terms lead to small dynamical changes in the horizon area coming
from typical quantum gravity properties. The first reason is quantum
uncertainty which does not allow a sharp condition $K_{\vp}=0$, and
the second space-discreteness and non-locality which implies that not
only $K_{\vp}$ at the horizon itself is relevant but also the values
in neighboring vertices which are not necessarily zero. For large
black holes, the correction terms are expected to be small:
uncertainty will not change the horizon area much compared to its
already large size, and in neighboring vertices of a semiclassical
state $K_{\vp}$ will still be extremely small. Thus, for large black
holes the horizon area is an excellent observable, while for
microscopic black holes large fluctuations are expected which may even
prevent horizons as they are known classically. This agrees with the
picture we have obtained from effective equations and matching
techniques before.

\section{Full theory}

The methods developed so far in symmetric models mimic those of the
full theory, with some adaptations to preserve the symmetry. In this
section, for completeness, we describe what this looks like in the
full theory and discuss applications which work without assuming
symmetries.

\subsection{Representation}

As discussed before, the full theory of loop quantum gravity is based
on holonomies for arbitrary edges in space and fluxes for surfaces,
forming the basic classical Poisson algebra. In the connection
representation, states are functionals on the infinite dimensional
space of connections \cite{DiffGeom,FuncInt,ALMMT} through holonomies,
and a dense subspace of the Hilbert space is spanned by cylindrical
functions
\begin{equation}
 \psi(A)=f_{\gamma}(h_{e_1}(A),\ldots,h_{e_n}(A))
\end{equation}
which depend on only finitely many holonomies. Since there is now no
symmetry requirement, the edges can be arbitrary curves in space and
form a graph $\gamma$ with vertices at their intersection points. The inner
product for two states associated with the same graph is given by
\begin{equation}
 \langle f_{\gamma}|g_{\gamma}\rangle=
  \int_{SU(2)^n} \prod_{e\in\gamma}{\rm d}\mu_H(h_e)
  f_{\gamma}(h_1,\ldots, h_n)^* g_{\gamma}(h_1,\ldots,h_n)
\end{equation}
with the Haar measure $\md\mu_H$ on the structure group SU(2). For two
functions with different graphs, they need to be extended to a bigger
one which is always possible by cutting edges or inserting new ones on
which the extended state depends trivially. An orthonormal basis is
given by spin network states \cite{RS:Spinnet,SpinNet}, associated
with graphs labeled by irreducible SU(2) representations $j_e$ at
edges and contraction matrices $C_v$ at vertices, of the form
\begin{equation}
 T_{\gamma,j,C}(A)=\prod_{v\in\gamma} C_v\cdot \prod_{e\in\gamma}
  \rho_{j_e}(h_e(A))
\end{equation}
where the representation matrices $\rho_{j_e}(h_e(A))$ evaluated in
edge holonomies are multiplied together in vertices according to the
symbols $C_v$.

Fluxes are quantized as derivative operators in the connection
representation since the densitized triad is conjugate to the Ashtekar
connection. Replacing the triad components in (\ref{Flux}) by
functional derivatives and acting on a cylindrical function, we obtain
\begin{eqnarray*}
\hat{F}_S f_{\gamma}&=& -8\pi i\gamma\hbar G \int_S {\rm d}^2y\tau^i n_a
\frac{\delta}{\delta A_a^i(y)} f_{\gamma}(h(A))\\
&=& -i\gamma\lP^2\sum_{e\in\gamma}\int_S {\rm d}^2y\tau^i n_a
\frac{\delta h_e}{\delta A_a^i(y)}\frac{\md f_{\gamma}(h)}{\md h_e}
\end{eqnarray*}
which has contributions only from intersection points $y$ of the
surface $S$ of the flux with the graph $\gamma$ associated with the
state. Moreover, each derivative operator for an intersection point
can be seen to be equivalent to an angular momentum operator such that
its spectrum is discrete and equidistant. Since there is a finite sum
over all such contributions, the spectrum of flux operators is
discrete, too. Not all the angular momentum operators involved
necessarily commute, and so triad operators do not always commute with
each other such that a triad representation does not exist
\cite{NonCommFlux} (unlike in the symmetric models studied before).

The densitized triad describes spatial geometry, and spatial quantum
geometry is encoded in flux operators. From the basic ones one can
construct geometrical operators such as the area \cite{AreaVol,Area}
or volume operator \cite{Vol2} which also have discrete spectra. Thus,
quantum spatial geometry is discrete in a precise way, given by the
spectra of geometric operators. The area spectrum is known completely,
but for the volume operator this is impossible to compute explicitly
since arbitrarily large matrices would have to be diagonalized.
Spatial geometry at the quantum level is thus rather complicated in
general if explicit calculations need to be done.

This translates to the Hamiltonian constraint and other operators, for
which the volume operator plays a crucial role. Classically, the
Hamiltonian constraint is given by \cite{AshVarReell}
\begin{equation}
 H[N] = (8\pi G)^{-1}\int\md^3x N(x) |\det
 E|^{-1/2}(F_{ab}^iE^a_jE^b_k
 \epsilon_{ijk}-2(1+\gamma^2)K_{[a}^iK_{b]}^jE_i^aE_b^k)
\end{equation}
with the curvature $F_{ab}^i$ of the Ashtekar connection, and the
extrinsic curvature $K_a^i=\gamma^{-1}(\Gamma_a^i-A_a^i)$ a function
of the basic variables through (\ref{GammaGen}). Both parts of the
constraint can be quantized using building blocks similar to
(\ref{HCont}), resulting in a well-defined operator \cite{QSDI} even
when matter Hamiltonians are included \cite{QSDV}. Edges for the
holonomies have to be chosen, which can be done in a diffeomorphism
invariant manner and even in such a way that the quantization is anomaly
free at least on states satisfying the diffeomorphism constraint
\cite{AnoFree}.

Which version of the quantization is the correct one, however, is
still an open issue since in particular the classical limit and that
of perturbations on a classical background (``gravitons'') are
difficult to analyze. Moreover, finding and interpreting solutions in
full generality is complicated by technical and conceptual problems.

It is thus important to devise approximation schemes, other than
symmetry reduction as employed before, in order to shed light on
physical properties of the full theory. One powerful possibility
consists in imposing an isolated horizon as a boundary \cite{IHPhase}
since boundary conditions imply that the constraint is not to be
imposed there (a constraint has lapse function going to zero at the
isolated horizon). Thus, also at the quantum level the constraint
operator can be ignored and aspects of the basic quantum
representation receive physical meaning. Indeed, boundary degrees of
freedom are obtained from intersections of spin network states with
the horizon surface, and flux operators are important to select
physical states corresponding to an isolated horizon. By counting
those states and comparing with the Bekenstein--Hawking expectation
the theory can be tested.

\subsection{Black hole entropy}

An isolated horizon $S$ with prescribed area $a_0$ as a boundary leads to
an additional boundary term in the symplectic structure \cite{IHPhase},
\begin{equation}
 \Omega=(8\pi\gamma G)^{-1}\int\md^3x \partial A_a^i\wedge \partial
 E^a_i+ \frac{a_0}{2\pi} (16\pi\gamma G)^{-1}\int_S \md^2y r_ir_j
 \partial A_a^i\wedge\partial A_b^j\epsilon^{ab}
\end{equation}
where we denote differentials on field space by $\partial$,
$\epsilon^{ab}$ is the anti-symmetric tensor on the boundary surface,
and $r_i$ the internal vector as in the definition of an isolated
horizon. The boundary term to the symplectic structure can be
recognized as that of a U(1) Chern--Simons theory which thus describes
the horizon degrees of freedom by the U(1) connection
$W_a=\frac{1}{2}r_iA_a^i$.

We quantize the full system by using quantum geometry in the bulk and
quantum Chern--Simons theory on the horizon. Doing this results in the
curvature $F=\md W$ becoming an operator with equidistant spectrum
which, via (\ref{IsoCond}) needs to be matched to the flux
$r_i\Sigma^i$ through the horizon. As shown before, quantum geometry
indeed implies a flux operator with equidistant spectrum such that the
matching is possible at the quantum level. Since also the pre-factors
match, there are always solutions to the horizon condition which can
now be counted, for a given area $a_0$, to compute the entropy as the
logarithm of the number of states.

This results in an expression for entropy which is proportional to the
horizon area \cite{ABCK:LoopEntro,IHEntro}, confirming expectations
from semiclassical considerations. Intuitively, entropy counts the
number of ways that one can construct a macroscopic horizon of area
$a_0$ from elementary discrete parts \cite{BekMuk} (which is
generalized in this picture since there is not just one elementary
type but different ones given by the spin label of an intersection
point with a spin network).  Since the discreteness scale is set by
the Barbero--Immirzi parameter $\gamma$, the number of possible such
configurations and thus entropy must depend on $\gamma$. Indeed,
$\gamma$ appears in the constant of proportionality between entropy
and area which allows us to fix $\gamma$ by requiring the
Bekenstein--Hawking law. Moreover, since there are different types of
black holes --- charged, distorted, rotating or with non-standard
matter couplings --- and the value is already fixed by the simplest
case of a Schwarzschild black hole, one can test the theory since now
entropy must result in the right way without any further parameter to
tune. This is indeed the case \cite{RotatingBH,NonminScalar},
providing a non-trivial test of the theory.

The scale of discreteness is then fixed which, since it must be small
enough, can already be confronted with observations. It turns out that
$\gamma=0.2735$ \cite{Gamma,Gamma2} is of the order one such that the
discreteness lies around $\sqrt{\gamma}\lP\approx \frac{1}{2}\lP$ and
is thus much too small to be observable directly. Indeed there have
long been reasons to expect a scale of discreteness around the Planck
length which is now confirmed by detailed calculations in loop quantum
gravity. It is not at all obvious that this comes about since there
are many non-trivial steps in the derivation, and mistakes in the
foundations of the theory could easily lead to larger values which
could already be in conflict with observations.

At this point it is important to consider the physical meaning of
$\gamma$. It can be seen as a fundamental parameter setting the scale
of discreteness which is thus characteristic of quantum gravity. (In
fact, one can express the continuum limit as a limit $\gamma\to0$
\cite{SemiClass}.) In usual arguments, this is expected to be done by
$\lP$, which has to appear anyway just for dimensional reasons.
However, in $\lP$ only the gravitational constant $G$ and Planck's
constant $\hbar$ enter such that the Planck length is already fixed by
classical gravity and quantum mechanics alone. Since these theories
are unrelated to full quantum gravity, there is no reason for $\gamma$
to equal one even though one can expect a value of the order one from
dimensional arguments. A precise value for the scale of discreteness
can only come from a detailed quantum theory of gravity and
calculations which are sensitive to the underlying discrete structure,
as realized by loop quantum gravity.

\section{Conclusion}

In the preceding sections we described the current status of what
black holes look like from the viewpoint of non-perturbative,
background independent quantum gravity. There are results obtained
with different approximations to the full theory which provide a
consistent picture of black holes without pathologies or puzzles, such
as the singularity problem or the information loss paradox, perceived
earlier from general relativity alone or from combinations of classical
gravity and quantum field theory on a background.

The main type of approximation used here is that of a symmetry
reduction as often employed in classical or quantum physics. This
allows to study the background independent quantum dynamics and its
characteristic features in different explicit ways. Compared to the
full theory, there are several technical simplifications for instance
from a volume operator with explicitly known spectrum. But also at a
conceptual level, the interpretation of solutions or physical
situations is simplified.

Even though special properties of a given symmetric model, such as
simplifying coordinate or field transformations, have not been made
use of and essential ingredients have rather been modeled on the full
theory, the question arises what one could do without symmetry
assumptions. For a fair judgment one has to bear in mind that
background independence in quantum field theory is a new concept,
which is introduced non-perturbatively. There are hardly any
comparable results in other realistic quantum field theories, and
quantum gravity introduces its own conceptual issues to the theory.
Moreover, the fact that common perturbative approximation schemes are
not available is a consequence of the property of gravity that a split
into a free field theory plus perturbations is not possible. One thus
has to deal with the fully non-linear framework which otherwise is
usually avoided in quantum field theory. Loop quantum gravity provides
a framework in which these hard questions, which sooner or later will
have to be faced by any approach to quantum gravity, are being
confronted directly.

A consequence of the non-linearity is that operators, even if they can
be defined in a well-defined manner, are by no means unique since
there are often factor ordering or regularization choices. Loop
quantum gravity, nevertheless, has succeeded in finding characteristic
effects from a background independent quantization. Details, of
course, depend on several quantization choices, but one can directly
investigate the robustness of results to ambiguities. As described
here, this allows one to solve conceptual problems in the physics of black
holes, and also in cosmology as detailed elsewhere
\cite{LoopCosRev,ICGC,LivRev}, while parameters can be fixed in detail by
consistency conditions or phenomenology.

\section*{Acknowledgements}

The author is grateful to Abhay Ashtekar, Rituparno Goswami, Roy
Maartens, Parampreet Singh and Rafal Swiderski for collaboration on
some of the results described here, and to Tom Roman for pointers to
references.

\newpage

\addcontentsline{toc}{section}{References}

\newcommand{\semicolon}{;}

\end{document}